\documentclass[fleqn,dvipdfmx]{emulateapj-rtx4}

\usepackage{graphicx}
\usepackage{amsmath}
\usepackage{bm}
\usepackage{color}

\newcommand{\jcop}{Journal of Computational Physics}
\newcommand{\sci}{Science}

\newcommand{\gcm}{\mbox{g~cm}^{-3}}
\newcommand{\cms}{\mbox{cm s}^{-1}}
\newcommand{\ergs}{\mbox{erg s}^{-1}}

\newcommand{\mdlx}[1]{$#1$M}

\newcommand{\tta}{t_{3\alpha}}
\newcommand{\ttar}{t_{3\alpha, {\rm r}}}
\newcommand{\ttas}{t_{3\alpha, {\rm s}}}
\newcommand{\tcc}{t_{\rm cc}}
\newcommand{\tccr}{t_{\rm cc,r}}
\newcommand{\tccs}{t_{\rm cc,s}}
\newcommand{\tdy}{t_{\rm dyn}}
\newcommand{\deta}{\epsilon_{3\alpha}}
\newcommand{\decc}{\epsilon_{\rm cc}}
\newcommand{\fsni}{$^{56}$Ni }
\newcommand{\fhe}{f_{\rm He}}

\renewcommand*{\thefootnote}{\fnsymbol{footnote}}

\begin{document}

\title{Hydrodynamical evolution of merging carbon-oxygen white dwarfs:
  their pre-supernova structure and observational counterparts}

\author{Ataru Tanikawa\altaffilmark{1,2}, Naohito
  Nakasato\altaffilmark{2}, Yushi Sato\altaffilmark{3,4}, Ken'ichi
  Nomoto\altaffilmark{5}\footnote{Hamamatsu Professor}, Keiichi
  Maeda\altaffilmark{6,5}, and Izumi Hachisu\altaffilmark{3}}
\affil{$^{1}$RIKEN Advanced Institute for Computational Science,
  7--1--26, Minatojima-minami-machi, Chuo-ku, Kobe, Hyogo, 650--0047,
  Japan \\ $^{2}$Department of Computer Science and Engineering,
  University of Aizu, Tsuruga Ikki-machi Aizu-Wakamatsu, Fukushima,
  965-8580, Japan \\ $^{3}$Department of Astronomy, Graduate School of
  Science, The University of Tokyo, 7--3--1, Hongo, Bunkyo-ku, Tokyo,
  113--0033, Japan \\ $^{4}$Department of Earth Science and Astronomy,
  College of Arts and Sciences, The University of Tokyo, 3--8--1,
  Komaba, Meguro-ku, Tokyo 153--8902, Japan \\ $^{5}$Kavli Institute
  for the Physics and Mathematics of the Universe (WPI), The
  University of Tokyo, 5--1--5, Kashiwanoha, Kashiwa, 277--8583, Japan
  \\ $^{6}$Department of Astronomy, Kyoto University,
  Kitashirakawa-Oiwake-cho, Sakyo-ku, Kyoto, 606--8502, Japan}

\begin{abstract}
  We perform smoothed particle hydrodynamics (SPH) simulations for
  merging binary carbon-oxygen (CO) white dwarfs (WDs) with masses of
  $1.1$ and $1.0$ $M_\odot$, until the merger remnant reaches a
  dynamically steady state. Using these results, we assess whether the
  binary could induce a thermonuclear explosion, and whether the
  explosion could be observed as a type Ia supernova (SN~Ia). We
  investigate three explosion mechanisms: a helium-ignition following
  the dynamical merger (`helium-ignited violent merger model'), a
  carbon-ignition (`carbon-ignited violent merger model'), and an
  explosion following the formation of the Chandrasekhar mass WD
  (`Chandrasekhar mass model'). An explosion of the helium-ignited
  violent merger model is possible, while we predict that the
  resulting SN ejecta are highly asymmetric since its companion star
  is fully intact at the time of the explosion. The carbon-ignited
  violent merger model can also lead to an explosion. However, the
  envelope of the exploding WD spreads out to $\sim 0.1R_\odot$; it is
  much larger than that inferred for SN~2011fe ($< 0.1R_\odot $) while
  much smaller than that for SN~2014J ($\sim 1R_\odot$). For the
  particular combination of the WD masses studied in this work, the
  Chandrasekhar mass model is not successful to lead to an SN~Ia
  explosion. Besides these assessments, we investigate the evolution
  of unbound materials ejected through the merging process (`merger
  ejecta'), assuming a case where the SN~Ia explosion is not triggered
  by the helium- or carbon-ignition during the merger. The merger
  ejecta interact with the surrounding interstellar medium, and form a
  shell. The shell has a bolometric luminosity of more than $2 \times
  10^{35}~\ergs$ lasting for $\sim 2 \times 10^4$~yr. If this is the
  case, Milky Way should harbor about $10$ such shells at any given
  time. The detection of the shell(s) therefore can rule out the
  helium-ignited and carbon-ignited violent merger models as major
  paths to SN~Ia explosions.
\end{abstract}

\keywords{binaries: close --- galaxies: evolution --- supernovae: general
--- white dwarfs --- hydrodynamics}

\section{Introduction}
\label{sec:intro}

The type Ia supernova (SN~Ia) is one of the brightest events in the
universe, and plays an important role as a cosmological distance
indicator. It is widely accepted that an SN~Ia is a thermonuclear
explosion of a carbon-oxygen (CO) white dwarf (WD), and that the
explosion is triggered by interaction between the CO~WD and its
companion star \citep{Nomoto84,Hillebrandt00}. However, it is still
controversial whether the companion star is a non-degenerate star
(single degenerate scenario; SD) \citep{Whelan73,Nomoto82}, or a
degenerate star (double degenerate scenario; DD)
\citep{Iben84,Webbink84}. There are other scenarios, for example, core
degenerate (CD) scenario in which the companion is an asymptotic giant
branch core \citep{Kashi11}, and collisional DD scenario in which two
CO~WDs collide in a dense stellar cluster, or in a multiple stellar
system \citep{Aznar-Siguan13,Aznar-Siguan14}.

The SD scenario has been well tested by recent observations
\citep[see][for a review]{Maoz14}. There are multiple observational
indications, some of which are for the SD scenario and the others are
against it, for different objects.  The observational studies against
the SD scenario include the following: \cite{Li11} have detected no
red giant star in the deep pre-explosion images of the site of
SN~2011fe. \cite{Schaefer12} have reported that no main sequence or
red giant stars are observed at the central region of an SN~Ia
remnant, SNR~0509-67.5\footnote{However, we should note that the null
  detection can be explained by spin-up/spin-down models, where the
  companion star evolved to become a helium WD during the spin-down
  phase of the CO~WD before the delayed carbon ignition occurs in the
  center \citep{DiStefano11,Justham11,Hachisu12}.}.  On the other
hand, the observations supporting the SD scenario include the
following: \cite{Dilday12} have observed SN Ia PTF~11kx, and have
found the evidence of a strong interaction between the SN ejecta and
circumstellar matter (CSM). Here, the CSM is thought to originate from
a symbiotic nova, which consists of a WD and a red giant
star.\footnote{For CSM in the CD scenario, see \cite{Soker13}.}

As described above, the SD scenario has been directly tested by many
studies. On the other hand, most of the `observational' support for
the DD scenario indeed come from `non-detection' -- it is supported
since the SD scenario is ruled out for particular
objects. \footnote{See, however, the earlier footnote for the SD
  scenario and \cite{Soker14} for the CD scenario of SN~2011fe.} It is
therefore necessary to assess the DD scenario directly, based on
theoretical predictions of what should be observed if the DD scenario
is the case. For this purpose, the following two questions should be
answered: (1) Whether a CO~WD which accretes materials from a
companion CO~WD results in a thermonuclear explosion. (2) Whether such
an explosion is observed as an SN~Ia.

Many previous studies have focused on the first question, finding many
possible paths in which a primary CO~WD could explode. These models
can be generally divided into the Chandrasekhar mass and
sub-Chandrasekhar mass models.  In the Chandrasekhar mass model, the
CO~WD reaches the central density higher than a critical density to
ignite explosive carbon burning, whereas it is not the case for the
sub-Chandrasekhar mass model.

In the Chandrasekhar mass model, two CO~WDs merge, and the merger
remnant evolves hydrostatically after the merger toward an explosion.
Whether the merger remnant explodes as an SN Ia depends on the
structure of the merger remnant. This structure has been intensively
investigated by means of two different numerical schemes: smoothed
particle hydrodynamics (SPH) simulations
\citep{Benz90,Guerrero04,Yoon07,LorenAguilar09,Raskin12} and
mesh-based hydrodynamics simulations \citep{DSouza06,Motl07}. Among
these works, \cite{Yoon07} have found a path to the SN Ia
explosion. Moreover, \cite{Zhu13} and \cite{Dan14} have performed
large parameter surveys for various binary CO~WD parameters, and have
searched for the systems which can explode as an SN Ia.

The sub-Chandrasekhar mass model can be subdivided into several
categories. First, \cite{Pakmor10} have suggested a `carbon-ignited
violent merger model' \cite[see
  also][]{Pakmor11,Pakmor12a,Pakmor12b}. In this model, hotspots
appear in the course of the merger of binary CO~WDs, and generate
carbon detonation leading to an explosion. Second, \cite{Pakmor13}
have also propounded a `helium-ignited violent merger model'. In this
model, a helium layer accreted onto a primary CO~WD from a companion
WD rises in temperature. It is suggested that the helium detonation
occurs at some point in this layer, and then the shock compression
triggers the carbon detonation inside the primary CO~WD, leading to an
explosion. This is an analog to the double detonation model
\cite[e.g.][]{Woosley80,Nomoto80,Nomoto82}, except for the nature of
the helium donor. Finally, \cite{Shen12}, \cite{Schwab12}, and
\cite{Ji13} have argued that binary CO~WDs could explode shortly after
the merger due to magnetohydrodynamical effects.

In this paper, we investigate the helium-ignited violent merger,
carbon-ignited violent merger, and Chandrasekhar mass models from two
points of view: (1) The first point is whether these models lead to a
successful ignition to initiate an explosion. For this purpose, we
perform SPH simulations of a merger of binary CO~WDs. However, we can
not directly follow initiation of its explosion; in order to follow
the initiation, we need SPH simulation with impossibly high space
resolution, say $1$~cm. Instead, we judge success or failure in the
explosion from the density and temperature obtained through the SPH
simulation. Especially, to investigate various possibilities in the
ignition process, we adopt the following strategy: Even when we infer
the success of a particular mode of the ignition, we do not stop our
simulation, and the SPH simulation results from the subsequent
evolution are used to test another mode of the ignition. This is
because our inference for a particular ignition mode is not decisive,
and because we want to test several models with different ignition
conditions. As described above, we use temperature to infer the
success or failure. However, temperature is vulnerable to random
noises in SPH simulations.  In order to obtain robust inference, we
carefully treat temperature in the following two ways. First, we adopt
two types of temperatures: raw and smoothed temperatures (described in
detail later). Second, we do not solve nuclear reactions, which are
sensitive to random noises, and could increase temperature in an
unstable manner.

(2) The second point is whether the expected outcome of the explosion
is consistent with observations of SNe~Ia. For this purpose, we adopt
several observational indications. The first test is the progenitor
radii of SN~2011fe and SN~2014J. The former and latter radii are
inferred to be less than $0.1R_{\odot}$
\citep{Nugent11,Bloom12,Zheng13,Mazzali14} and more than $1R_{\odot}$
\citep{Goobar14}, respectively. As another test, we discuss $^{56}$Ni
distribution and the ejecta geometry, both of which are thought not to
be highly asymmetric for nearby SNe \citep{Maund13,Soker14} and for SN
remnants.

We mainly investigate a merger of a binary consisting of two CO~WDs
with masses of $1.1$ and $1.0$ $M_{\odot}$. For a benchmark, we also
follow evolution of a binary with masses of $0.9M_{\odot}$ CO~WD and
$0.6M_{\odot}$ CO~WD. Other combinations are investigated elsewhere
\citep{Sato15}.

In sum, we find that for the particular binary parameters studied in
this paper, the helium- and carbon-ignited violent merger models lead
to an explosion, while the Chandrasekhar mass model does not,
according to our inference. However, as we mentioned above, our
inference is not decisive due to various uncertainties. We therefore
suggest a way of evaluating our inference and constrain the fates of
WD mergers, based on the insights obtained through the SPH
simulations. If the system indeed does not immediately lead to an SN
Ia explosion, the binary merger should leave its merger WD remnant and
the unbound materials ejected from the system during the dynamical
phase of the merger process (hereafter `merger ejecta'). The merger
ejecta interact with interstellar medium (ISM), and form a shell
(hereafter 'merger shell'), which is analogous to formation of an SN
remnant. If we find these merger remnants and merger shells, we can
dismiss our inference and alternatively we can use these observational
counterparts against the violent merger scenarios leading to an SN
Ia. We discuss the detectability of such events, in particular the
merger shell. We thereby suggest that it is possible to detect such
events.

This discussion has another benefit. The merger shell can be detected
not only from pre-explosion images of a site of an SN~Ia, but also in
the post-explosion observations, if the SN~Ia happens in the
Chandrasekhar mass model\footnote{The Chandrasekhar mass model can be
  successful when binary CO~WDs consist of the mass combinations
  different from the one studied in this paper
  \citep[see][]{Sato15}.}. The detection of the merger shell can
therefore directly support the Chandrasekhar mass model as a result of
the WD merger.

This paper is structured as follows. In section~\ref{sec:method}, we
describe the methods of our simulations. In section~\ref{sec:results},
we show the results from our simulations. In
section~\ref{sec:verification}, we assess whether our binary CO~WDs
can explode, and whether the explosion can be observed as an SN~Ia. In
section~\ref{sec:detectability}, we discuss the detectability of the
merger remnant and merger shell. Finally, we summarize our findings in
section~\ref{sec:summary}.

\section{Method}
\label{sec:method}

In this section, we describe the methods of our simulations. In
section~\ref{sec:sph}, we briefly present schemes in our SPH
simulations. In section~\ref{sec:define}, we define several quantities
used throughout the paper. In section~\ref{sec:setup}, we show how to
set up an initial condition of binary CO~WDs. In
section~\ref{sec:run}, we summarize a set of physical and numerical
parameters used in this study.  In section~\ref{sec:environment}, we
introduce our computing environment.

\subsection{SPH simulation}
\label{sec:sph}

We solve Lagrangian hydrodynamics equations with self gravity by means
of SPH simulations. Each SPH particle is evolved by the following
equations:
\begin{align}
  \dot{\bm{v}}_i &= - \frac{\nabla P_i}{\rho_i} +
  \bm{g}_i, \label{eq:eom} \\ 
  \dot{u}_i &= - \frac{P_i}{\rho_i} \left( \nabla \cdot \bm{v}_i
  \right), \label{eq:eou}
\end{align}
where $\bm{v}_i$, $u_i$, $P_i$, and $\rho_i$ are the velocity,
specific internal energy, pressure, and (mass) density of
$i$-particle, respectively, and $\bm{g}_i$ is gravity exerting on
$i$-particle. The over-dots indicate the first time derivative. The
nabla symbol means an operator of $(\partial/\partial x,
\partial/\partial y, \partial/\partial z )$.

We briefly explain our SPH formulations. In our SPH simulations, we
solve the `vanilla ice' SPH equations. We adopt a cubic spline kernel
for the SPH kernel interpolation. The SPH kernel is modified in the
same way as \cite{Thomas92}. Similarly to \cite{Rosswog00}, we adopt
the treatment of time-dependent artificial viscosity \citep{Morris97},
combined with a recipe which suppresses the viscosity from shear
motion \citep{Balsara95}. This is described in detail in
appendix~\ref{sec:viscosity}. We set the length of the SPH kernel of a
particle, such that the arithmetic average number of neighbor
particles over all the particles, $\ensuremath{\langle n_i \rangle}$,
is $150$. Hereafter, ``$\ensuremath{\langle \rangle}$'' indicates an
arithmetic average of quantities over all the particles. Neighbor
particles of $i$-particle, $n_i$, are defined as particles whose
distances from $i$-particle is less than the kernel length of
$i$-particle.

We use an equation of state (EoS) as functions of $\rho_i$ and $u_i$
in order to get not only $P_i$, but also raw temperature $T_i$ and
sound speed $c_{{\rm s},i}$. For the EoS, we adopt the Helmholtz EoS
\citep{Timmes00}. This EoS includes thermal radiations, an ideal gas
of ions, an electron-positron gas with an arbitrary degree of
relativity and degeneracy. The EoS requires chemical compositions of
fluids. We assume that the chemical composition is uniformly fixed to
$50$ percent of carbon, and $50$ percent of oxygen in the number
fraction. The chemical composition is fixed during the whole
simulations, since we do not consider nuclear reactions.

The gravity is calculated as follows. The gravity $\bm{g}_i$ is the
sum of the Newtonian gravity on $i$-particle exerted by all the other
particles. We introduce so-called Plummer softening to the Newtonian
gravity. Thus, it is expressed as
\begin{align}
  \bm{g}_i = \sum_{j \neq i} Gm_j\frac{\bm{r}_j -
    \bm{r}_i}{\left(|\bm{r}_j - \bm{r}_i|^2 + \varepsilon^2
    \right)^{3/2}},
\end{align}
where $m_j$ and $\bm{r}_j$ are respectively the mass and position
vector of $j$-particle, $G$ is the gravitational constant, and
$\varepsilon$ is the gravitational softening, fixed to $\varepsilon =
3 \times 10^6$~cm. In practice, we calculate the gravity with an
octree algorithm \citep[e.g.][]{Barnes86}. In such algorithms, gravity
exerted on $i$-particle by distant particles is approximated as its
multipole moment. We consider the multipole moment up to dipole
moment. In order to define whether particles are distant or not, we
use Multipole Acceptance Condition (MAC). We choose the same MAC as
introduced by \cite{Salmon94}. The MAC has one accuracy parameter,
$\Delta$, which is the same notation as in \cite{Nakasato12}. The
accuracy parameter $\Delta$ has the dimension of mass divided by
square of length. As $\Delta$ becomes smaller, gravity is calculated
with a higher degree of accuracy. At the beginning of our simulation,
we determine $\Delta$ from $\bm{g}_i$ at the initial time, such as
\begin{align}
  \Delta = \eta \frac{\ensuremath{\langle |\bm{g}_i|
      \rangle}}{G}, \label{eq:mac}
\end{align}
where a relative error of gravity on a particle is ensured to be less
than $\eta$. Except for $t=0$, we adopt $\Delta=1.7 \times
10^{14}$~gcm$^{-2}$, which corresponds to $\eta = 0.01$. At $t=0$, we
adopt $\Delta \sim 1.0 \times 10^9$~gcm$^{-2}$. The reason why
$\Delta$ at $t=0$ is extremely smaller than $\Delta$ at $t > 0$ is as
follows. We do not know the gravity $\bm{g}_i$ at $t=0$. Therefore, we
do not know what $\Delta$ corresponds to $\eta=0.01$ at $t=0$. In
order to avoid obtaining the gravity with a relative error larger than
$0.01$, we use sufficiently small $\Delta$ at $t=0$. As a result of
the calculation of the gravity at $t=0$, it becomes clear that $\Delta
\sim 1.0 \times 10^9$~gcm$^{-2}$ corresponds to $\eta = 6 \times
10^{-8}$. In other words, we calculate the gravity with a relative
error much less than $0.01$ at $t=0$.

\subsection{Definitions}
\label{sec:define}

In this section, we define several quantities used throughout this
paper.

We define two types of temperatures: raw and smoothed
temperatures. The raw temperature of a particle, $T_{{\rm r},i}$, is
directly obtained from the Helmholtz EoS as a function of $\rho_i$ and
$u_i$, as described in section~\ref{sec:sph}. Using $T_{{\rm r},i}$,
the smoothed temperature of a particle, $T_{{\rm s},i}$, is calculated
as
\begin{align}
  T_{{\rm s},i} = \sum_j^{N} T_{{\rm r},j} \frac{m_j}{\rho_j}
  W(|\bm{r}_j-\bm{r}_i|,h_j),
\end{align}
where $W$ is the SPH kernel. This smoothed temperature is similar to
``SPH-smoothed temperature'' in \cite{Dan14}.

We define a timescale of a nuclear reaction, $t_{\rm nuc}$, as
\begin{align}
  t_{\rm nuc} = \frac{c_{\rm p}T}{\epsilon_{\rm nuc}}, \label{eq:tta}
\end{align}
where $c_{\rm p}$ is specific heat at constant pressure,
$\epsilon_{\rm nuc}$ is an energy generation rate per unit mass for
the nuclear reaction, and $T$ is either the raw or smoothed
temperature. We indicate $t_{\rm nuc,r}$ and $t_{\rm nuc,s}$ as the
timescale of a nuclear reaction, when we calculate the timescale with
the raw and smoothed temperatures, respectively. We replace the
subscripts ``nuc'' in equation~(\ref{eq:tta}) with ``$3\alpha$'' for
the triple-alpha reaction, and ``cc'' for the $^{12}$C + $^{12}$C
reaction. We calculate an energy generation rate by the triple-alpha
reaction in the same way as \cite{Kippenhahn90}:
\begin{align}
  \deta = q_{\rm he} f_{3\alpha} \rho^2 X_4^3 T_8^{-3} \times \exp
  \left( -44.027 / T_8 \right),
\end{align}
where $q_{\rm he}=5.09 \times 10^{11}$~[erg~g$^{-1}$~s$^{-1}$],
$f_{3\alpha} = \exp(2.76 \times 10^{-3} \rho^{1/2} T_8^{-3/2})$ is the
weak electron screening factor \citep{Salpeter54,Clayton68}, $X_4$ is
the mass fraction of helium, and $T_8 = T/10^8$, which is the same
choice as \cite{Dan14}. We also define an energy generation rate by
$^{12}$C + $^{12}$C as
\begin{align}
  \decc = \rho q_{\rm c} Y_{\rm C}^2 A_{\rm T9} \exp(- Q / T_{\rm
    9a}^{1/3} + f_{\rm cc}),
\end{align}
where $q_{\rm c}=4.48 \times 10^{18}$ [erg~mol$^{-1}$], $f_{\rm cc}$
is a screening factor \citep{Blinnikov87}, $A_{\rm T9}=8.54 \times
10^{26} T_{\rm 9a}^{5/6} T_{\rm 9}^{-3/2}$ [s$^{-1}$~cm$^3$],
$Q=84.165$, $T_{\rm 9}=T/(10^9$K), and $T_{\rm 9a}=T_{\rm
  9}/(1+0.067T_9)$ \citep{Fowler75}. We calculate carbon abundance as
$Y_{\rm C}=n_{\rm C}/(\rho N_{\rm a}) = 0.033$ [g$^{-1}$], where
$n_{\rm C}$ is the number density of carbon, and $N_{\rm a}$ is
Avogadro constant.

We define the local dynamical timescale as
\begin{align}
  \tdy = \left( 24 \pi G \rho \right)^{-1/2}, \label{eq:tdyn}
\end{align}
which is the same choice as that of \cite{Nomoto82}.

We introduce a shock detector of a particle in order to search for
shock-heated regions. The shock detector is defined as
\begin{align}
  D_{{\rm s},i} &= - f_i \left[ \frac{h_i}{c_{{\rm s},i}} \left(
    \nabla \cdot \bm{v}_i \right) \right],
\end{align}
where $f_i$ is Balsara switch (see appendix~\ref{sec:viscosity}). This
indicates how strongly a fluid element is compressed. The element is
compressed when $D_{{\rm s},i} > 0$, and is extended when $D_{{\rm
    s},i}<0$. The critical value between shocked and unshocked regions
is about unity, but is not severe. Although it is just an indicator,
it should be useful to grasp where fluid elements are compressed by
shock waves.

We present the definition of merger ejecta as follows. We consider a
specific orbital energy of a particle at the time $t$ as $b_i = \phi_i
(t) + 0.5 |\bm{v}_i(t)|^2$, where $\phi_i(t)$ and $\bm{v}_i(t)$ are a
specific potential energy and velocity of the particle at the time
$t$. If a particle has $b_i(t)>0$, it is unbound at the time $t$. A
particle which keeps $b_i(t') \ge 0$ during $t' \ge t$ is defined as a
merger ejecta at the time $t$. This means that a particle unbound
temporarily is not counted as a merger ejecta. We also show the
definition of a terminal velocity of a merger ejecta:
\begin{align}
  v_{{\rm ej},i}(t) = \left( 2 b_i \right)^{1/2}. \label{eq:vej}
\end{align}
We can interpret the terminal velocity as a velocity of a particle at
infinity.

\subsection{Initial condition setup}
\label{sec:setup}

We set up an initial condition in our SPH simulations in three
steps. Our setup method is the same as that of \cite{Dan11}, unless
otherwise noted. In the first step, we generate two single CO~WDs
individually. In the second step, we choose particles from each of the
single CO~WDs, and assume that these particles consist of pure
helium. In the third step, we combine the two CO~WDs in the same
frame, and form binary CO~WDs. We explain these steps in detail in the
following.

We can divide the first step into five substeps. In the first step, we
make 1-dimensional density profile of a fully-degenerate CO~WD with
uniform temperature of $10^6$~K. In the second substep, we map
particles in such a way that their mass densities and specific
internal energies are consistent with the single CO~WD. In the third
substep, we relax these particles as follows. We evolve these
particles for $20$~s in the simulation time by means of SPH simulation
which is different from that described in section~\ref{sec:sph} in two
points. One is that a specific internal energy of each particle is
fixed. In other words, we do not solve equation~(\ref{eq:eou}). The
other is that each particle receives a damping force against its
motion. The damping force on $i$-particle is added to the right-hand
side of equation~(\ref{eq:eom}), and given by
\begin{align}
  \left({\dot{\bm{v}}_i}\right)_{\rm damp} = -
  \frac{\bm{v}_i}{\tau_{\rm damp}}, \label{eq:damp}
\end{align}
where we set $\tau_{\rm damp} = C_{\rm damp} \Delta t$, and $C_{\rm
  damp}=128$. We adopt $C_{\rm damp}=128$
experientially. Nevertheless, the value of $C_{\rm damp}$ does not
have the impact on the structure of the CO~WD. The structure of the
CO~WD in the case of $C_{\rm damp}=128$ is almost the same as that in
the case of $C_{\rm damp}=64$. In the fourth substep, we relax these
particles again by evolving them during $80$~s by means of SPH
simulation described in section~\ref{sec:sph}. The purpose of this
substep is to avoid mixing of the helium and carbon-oxygen
particles. This substep has not been done in \cite{Dan11}. In the
fifth substep, we shift the center of mass in the positions of the SPH
particles to the origin of coordinates, and the center of mass in the
velocities to zero.

\begin{figure*}
  \begin{center}
    \includegraphics[width=10.0cm]{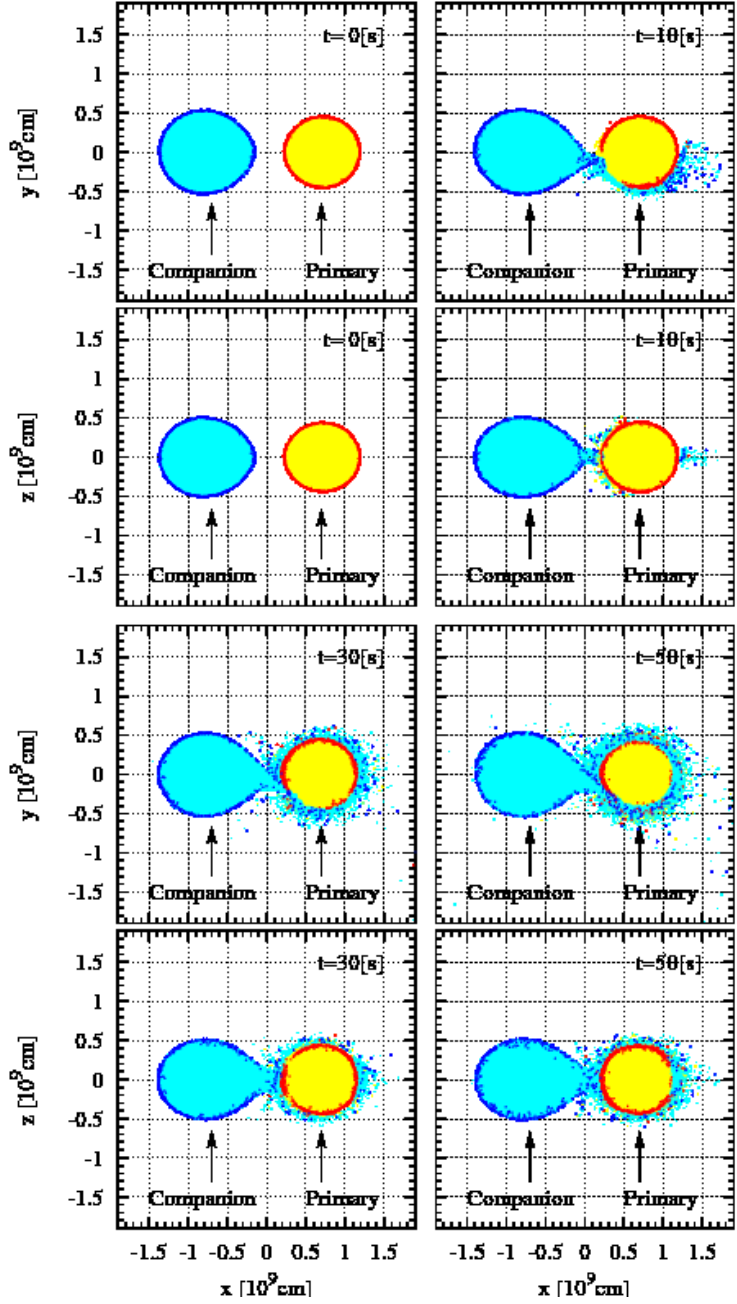}
  \end{center}
  \caption{Distribution of helium and carbon-oxygen particles at
    $t=0$, $10$, $30$, and $50$~s in model \mdlx{5.5}. The red and
    blue points indicate the helium particles of the primary and
    companion WDs, respectively. The yellow and light blue points show
    the carbon-oxygen particles of the primary and companion WDs,
    respectively.}
  \label{fig:init_helium}
\end{figure*}

In the second step, we choose the outermost particles of each of the
single CO~WDs, and assume these particles consist of pure helium. We
call these particles `helium particles'. We define a helium fraction,
$f_{\rm He}$, as the ratio of helium mass to CO mass in these
particles. We note that the EoS of these helium particles is {\it not}
helium one, but carbon-oxygen one, but it would not introduce a large
error for our purposes. We compare the radius of a pure CO~WD with
that of a CO~WD with helium ($f_{\rm He} = 1 \times 10^{-3}$, $3
\times 10^{-4}$, and $4 \times 10^{-5}$) by performing 1-dimensional
hydrostatic calculation, and find that the difference among their
radii is at most $1$ percent. Proactively, we show that the helium
particles are not mixed carbon-oxygen ones without merging.
Figure~\ref{fig:init_helium} indicates the distribution of the helium
and carbon-oxygen particles. Particularly, as seen in the companion WD
which does not accrete materials, the helium particles keep staying on
the surface of the companion WD.

We can divide the third step into two substeps. In the first substep,
we put two single CO~WDs in the same frame. Hereafter, we call the
more massive CO~WD `the primary', and the less massive one `the
companion'. We set a separation between the primary and companion,
such that the companion's Roche-lobe radius is $C_{\rm lobe}$ times
larger than the companion's radius. We estimate the Roche-lobe radius
from an approximate formula of \cite{Eggleton83}. We can express the
separation between the primary and companion, $a_0$, as
\begin{align}
  a_0 = C_{\rm lobe} R_{\rm c} \left[ \frac{0.49 q^{2/3}}{0.6 q^{2/3} +
      \log (1 + q^{1/3})} \right]^{-1},
\end{align}
where $R_{\rm c}$ is the companion's radius, and $q$ is the ratio of
the companion's mass ($M_{\rm c}$) to the primary's mass ($M_{\rm
  p})$. We set $C_{\rm lobe}=2$ in this study. Hereafter, we define
the separation between the primary and companion as the distance
between their centers of mass. They are on an circular orbit around
the origin of coordinates.  In the second substep, we relax the
configuration of particles composing the binary CO~WDs, and
simultaneously decay the orbit of the binary. For the relaxation, we
again introduce the damping force given by
equation~(\ref{eq:damp}). We decay their orbit at every $\Delta t_{\rm
  decay}$ seconds. The extent of the decay is given by
\begin{align}
  \Delta a_{\rm decay} = \frac{a}{\tau_{\rm decay}} \Delta t_{\rm
    decay},
\end{align}
where $a$ is the binary separation before the decay. We can write
$\tau_{\rm decay}$ as
\begin{align}
  \tau_{\rm decay} = \frac{1}{C_{\rm decay}}
  \frac{1}{\sqrt{G\rho_{c,0}}},
\end{align}
where $\rho_{\rm c,0}$ is the overall mass density of the companion
before this relaxation and orbital decay, and written as $\rho_{\rm
  c,0}=M_{\rm c}/(4\pi R_{\rm c}^3/3)$. We adopt $\Delta t_{\rm
  decay}=1/64$~s and $C_{\rm decay}=0.05$. In this relaxation and
orbital decay, we take a co-rotating frame of reference. Therefore, we
add centrifugal and Coriolis terms to equation~(\ref{eq:eom}),
expressed as
\begin{align}
  \left( \dot{\bm{v}} \right)_{\rm corotate} = - \bm{\omega} \times
  (\bm{\omega} \times \bm{r}_i) - 2 \bm{\omega} \times \bm{v}_i,
\end{align}
where $\bm{\omega}$ is the angular velocity vector, and
$|\bm{\omega}|=[G(M_{\rm p}+M_{\rm c})/a^3]^{1/2}$. We stop this
process, $\Delta t_{\rm decay}$ after their separation is decreased to
less than a critical separation $a_{\rm crit}$. This is different from
the setup method of \cite{Dan11}, who stop this process when any
particle exceeds the Roche lobe.

\begin{figure*}
  \begin{center}
    \includegraphics[width=16.0cm]{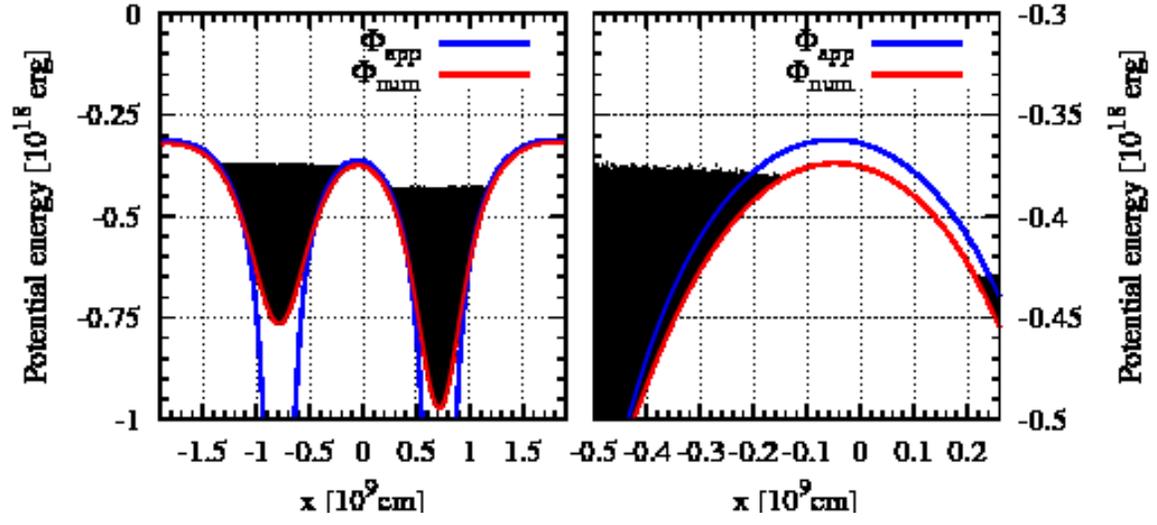}
  \end{center}
  \caption{Potential energies of particles as a function of their
    $x$-coordinate. The definition of $\Phi_{\rm app}$ and $\Phi_{\rm
      num}$ are in the main text.}
  \label{fig:init_carbonoxygen}
\end{figure*}

In Figure~\ref{fig:init_carbonoxygen}, black dots show the potential
energies of particles in the frame corotating with the binary. This
figure corresponds to figure~3 of \cite{Dan11}. The blue curves
($\Phi_{\rm app}$) indicate the potential energies of the field, where
the primary and companion are approximated to be point mass at their
center of mass. The potential energies are expressed as
\begin{align}
  \Phi_{\rm app} = - \frac{GM_{\rm p}}{|\bm{r}-\bm{r}_{\rm p}|} -
  \frac{GM_{\rm c}}{|\bm{r}-\bm{r}_{\rm c}|} - \frac{1}{2} \left(
  \bm{\omega} \times \bm{r} \right)^2,
\end{align}
where $\bm{r}_{\rm p}$ and $\bm{r}_{\rm c}$ are the position of the
center of mass of the primary and companion, respectively. The red
curves ($\Phi_{\rm num}$) indicate the potential energies of the
field, where the potential energies among particles are calculated
numerically. The potential energies are expressed as
\begin{align}
  \Phi_{\rm app} = \Phi(\bm{r}) - \frac{1}{2} \left( \bm{\omega}
  \times \bm{r} \right)^2,
\end{align}
where $\Phi(\bm{r})$ is the potential energies among particles. The
binary is tidally locked.

\subsection{Simulation run}
\label{sec:run}

We simulate mergers of two types of binary CO~WDs: pairs of
$1.1M_{\odot}$ and $1.0M_{\odot}$ CO~WDs, and $0.9M_{\odot}$ and
$0.6M_{\odot}$ CO~WDs. The latter is used for a benchmark, since such
a pair of CO~WDs has been widely investigated in various studies, such
as \cite{Yoon07}.

We perform the simulations for binary CO~WDs with masses of
$1.1M_{\odot}$ and $1.0M_{\odot}$ as follows. We set a separation
between the primary and companion, such that $a_{\rm crit}=1.5 \times
10^9$~cm. We choose the helium fraction as $f_{\rm He}=4 \times
10^{-5}, 3 \times 10^{-4}$, and $1 \times 10^{-3}$ for both the
primary and companion. We adopt various mass resolutions, where the
numbers of SPH particles used to resolve $0.1 M_{\odot}$ are $64k$,
$128k$, $256k$, and $512k$ (where $1k=2^{10}=1024$).  The total
numbers of particles in each model are therefore about $1.4$, $2.8$,
$5.5$, and $11 \times 10^{6}$. We name these models `model $x$M'
($x=1.4, 2.8, 5.5$, and $11$) after the total number of particles used
in each run. We follow the evolutions of binary CO~WDs for $500$~s of
the simulation time, except for model \mdlx{11}. In all the models,
they merge after the binary components orbit around each other several
times. In all models but for model \mdlx{11}, the simulation is
followed until the merger remnants reach a dynamically steady
state. We stop the simulation of \mdlx{11} just after their merger,
since the simulation time is quite long.

For binary CO~WDs with masses of $0.9M_{\odot}$ and $0.6M_{\odot}$, we
separate the primary and companion by $a_{\rm crit}=2.45 \times
10^9$~cm.  The number of SPH particles to resolve $0.1 M_{\odot}$ is
$64k$; the total number of particles is about $980 \times 10^{3}$. We
evolve it for $1000$~s. At that time, it reaches a dynamically steady
state.

\subsection{Computing environment}
\label{sec:environment}

We use a code called OTOO \citep{Nakasato12}, which stands for `OcTree
On Opencl'. The OTOO code (hereafter, {\tt OTOO}) supports a variety
of astronomical particle simulations, such as $N$-body and SPH
simulations. {\tt OTOO} utilizes an octree algorithm
\citep[e.g.][]{Barnes86} for fast calculations of particle-particle
interactions. It is optimized to multi- and many-core architectures on
shared-memory environment. It is relatively machine-independent, since
it is implemented with OpenCL.

We perform each SPH simulation on a single node of a supercomputer
HA-PACS at Center for Computational Sciences, University of Tsukuba. A
single node of HA-PACS consists of two CPUs plus four GPUs. The CPUs
are Intel Sandy Bridge-EP-8, and the GPUs are NVIDIA Tesla M2090. In
this configuration, {\tt OTOO} spends $1.7$, $3.3$ $6.5$, and $13$~s
for every timestep for models \mdlx{1.4}, \mdlx{2.8}, \mdlx{5.5}, and
\mdlx{11}, respectively. The total wall-clock time for these models
are about $160$, $360$, $670$, and $1200$ hours.

\section{Simulation results}
\label{sec:results}

In this section, we show results of our simulations. In
section~\ref{sec:overview}, we overview the time evolution of the
binary CO~WDs. In section~\ref{sec:comparison}, we compare our results
with previous studies.

\subsection{Overview}
\label{sec:overview}

\begin{figure*}
  \begin{center}
    \includegraphics[width=16.0cm]{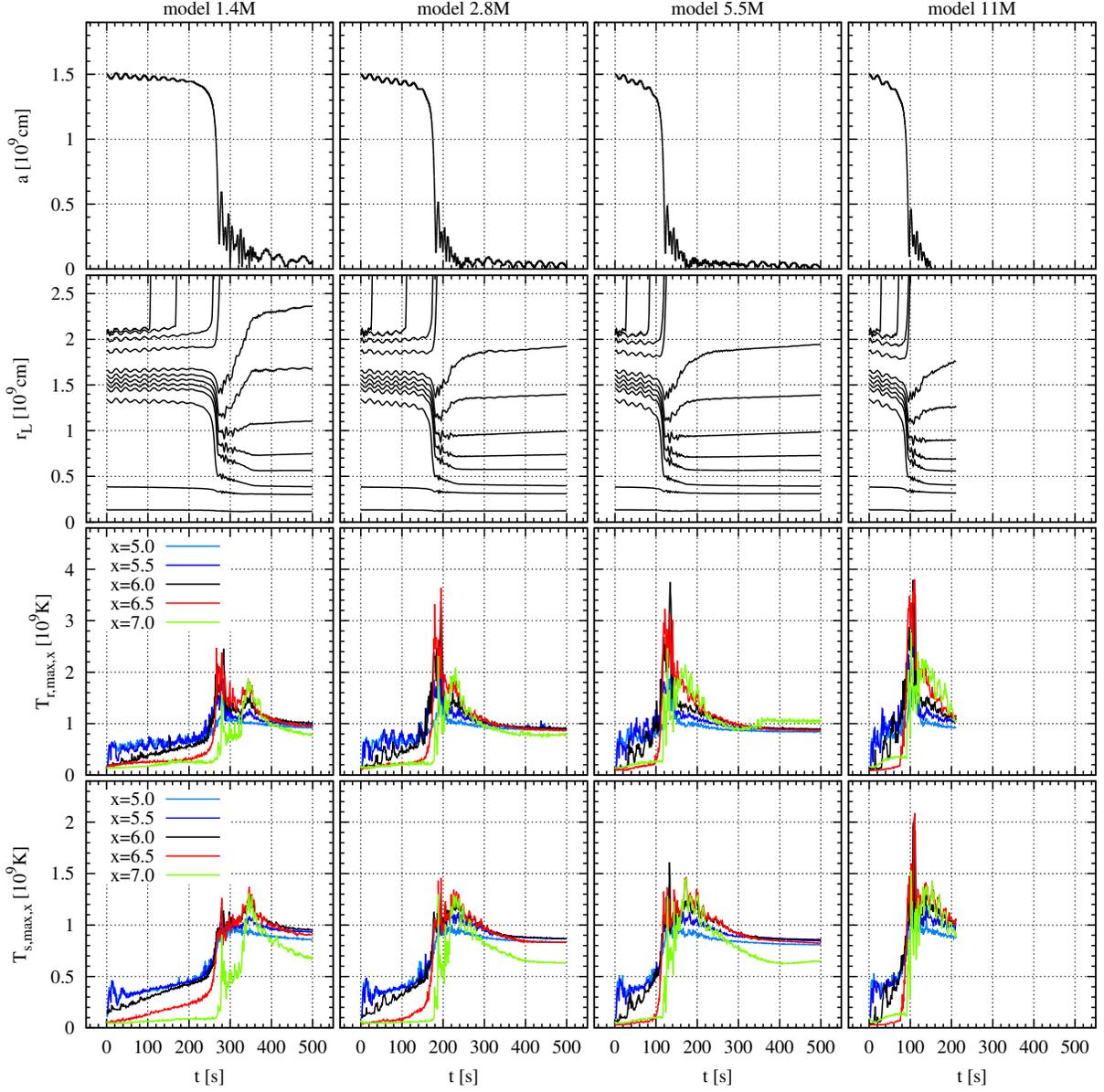}
  \end{center}
  \caption{Time evolution of binary CO~WDs in models \mdlx{1.4},
    \mdlx{2.8}, \mdlx{5.5}, and \mdlx{11} from left to right. Each top
    panel shows the separation between the primary and companion. Each
    second top panel draws $10$, $50$, $60$, $65$, $70$, $75$, $80$,
    $90$, $99$, $99.9$, $99.99$, and $99.999$ percent Lagrangian radii
    (defined in the main text) from bottom to top. The second bottom
    and bottom panels indicate the maximum of raw and smoothed
    temperatures in a range of mass density shown in the left panels.}
  \label{fig:hotspot_ov}
\end{figure*}

In Figure~\ref{fig:hotspot_ov}, we present the time evolution of the
binary CO~WDs in models \mdlx{1.4}, \mdlx{2.8}. \mdlx{5.5}, and
\mdlx{11}. In the top panels, we show the separations of the
binaries. In the second top panels, we present mass distribution of
the binaries. For this purpose, we use `$x$ percent Lagrangian radii'
defined as follows: Each of them is a radius of a sphere which
encloses $x$ percent of the total mass of the binary CO~WDs, and whose
center is the center of mass of the primary. In the second top panel,
we draw $10$, $50$, $60$, $65$, $70$, $75$, $80$, $90$, $99$, $99.9$,
$99.99$, and $99.999$ percent Lagrangian radii from bottom to top. In
the second bottom and bottom panels, we depict the maximum $T_{{\rm
    r},i}$ and $T_{{\rm s},i}$, respectively, in a range of mass
density of $10^{x} < \rho_i / (\gcm) < 10^{x+0.5}$. These maximum
$T_{{\rm r},i}$ and $T_{{\rm s},i}$ are, respectively, indicated as
$T_{{\rm r,max},x}$ and $T_{{\rm s,max},x}$. We sample all the above
quantities at every $1$~s.

We can see in the top panels that the separation is steeply decreased
at $t \sim 270$, $180$, $120$, and $100$~s for models \mdlx{1.4},
\mdlx{2.8}, \mdlx{5.5}, and \mdlx{11}, respectively. At these times,
the binaries merge. We call these times `merger times'. The binaries
rotate around each other at least $5$ times before the dynamical
merger, since their periods are about $22$~s. It is difficult to avoid
such different merger times for different resolutions, since the mass
transfer from the companion to the primary is quite unstable and
chaotic. The separation oscillates with the periods of the
binaries from the initial time to the merger time. This is due to
non-zero eccentricities of the orbits of these binaries.

In our simulation, the merger time is relatively smaller than previous
studies, such as those of \cite{Dan11}. This may be because we stop
the relaxation process of the binary when the binary separation
becomes smaller than a critical separation. We may make the binary
separation too small. In section~\ref{sec:verification}, we discuss
this effect on whether the thermonuclear explosion becomes successful
or not.

We follow the time evolution of the inner mass distribution ($\le$ 90
percent Lagrangian radii) of model \mdlx{5.5} as an example. This is
instructive, since the mass distribution in other models is similar to
that in model \mdlx{5.5}, except their merger times. In model
\mdlx{5.5}, the $10$ and $50$ percent Lagrangian radii are not changed
throughout the simulation. These radii represent the primary's
material. This means that the merger has little effect on the mass
distribution of the primary. On the other hand, the $60$ -- $90$
percent Lagrangian radii are drastically decreased around the merger
time ($t \sim 120$~s). These radii contain the companion's mass. This
means that most of the companion's mass is accreted to the primary at
once at the merger time. During about $50$~s after the merger time,
the mass distribution of the accreted material is still
changing. During this time, the $60$, $65$, and $70$ percent
Lagrangian radii decrease gradually, whereas the $80$, and $90$
percent Lagrangian radii expand rapidly. After this time, the
materials below the $70$ percent Lagrangian radii keep constant in
their radii, while the materials outer than this in the mass
coordinate expand slowly.  We conclude that the mass distribution, in
particular the interior of the `merger remnant', is dynamically steady
at the end of our simulation ($t=500$~s).

Below, the evolution of the outer mass distribution, above than $90$
percent Lagrangian radii., is explained for model \mdlx{5.5}. The
$99.99$ and $99.999$ percent Lagrangian radii rapidly increase well in
advance of the merger time. Similarly, the $99$ and $99.9$ percent
Lagrangian radii rapidly increase around the merger time. The
evolution of these Lagrangian radii means that a substantial amount of
materials is ejected before/at the merging process, which surround the
binary system. This matter can/should affect observations of outcome
of the merger, e.g., an explosion, as discussed in detail in
section~\ref{sec:verification}.

The highest raw temperature is achieved for $20$~s following the
merger time in each model. Hereafter, this peak of temperature is
called `first peak'. The temperature at the first peak is $2.4 \times
10^9$~K, $3.6 \times 10^9$~K, $3.8 \times 10^9$~K, and $3.8 \times
10^9$~K for models \mdlx{1.4}, \mdlx{2.8}, \mdlx{5.5}, and \mdlx{11},
respectively. This temperature seems to converge to $3.8 \times
10^9$~K toward the higher resolution. The smoothed temperature also
reaches a high value around the first peak. In fact, the smoothed
temperature at the first peak is the highest among the values obtained
for the whole evolution of the system, in each of models \mdlx{2.8},
\mdlx{5.5}, and \mdlx{11}. It is $1.4 \times 10^9$~K, $1.6 \times
10^9$~K, and $2.1 \times 10^9$~K for models \mdlx{2.8}, \mdlx{5.5},
and \mdlx{11}, respectively. This temperature however does not
converge even with the highest resolution among our runs. Moreover,
this smoothed temperature is much lower than the raw temperature at
the first peak. If the mass resolution becomes extremely high, the
smoothed temperature at the first peak may converge, and may become
consistent with the raw temperature at the first peak. We do not
discuss the convergence of the smoothed temperature anymore in this
paper, but we keep in mind that we may underestimate temperature when
we adopt the smoothed temperature.

Both of the raw and smoothed temperatures in each model have another
peak about $20$~s after the first peak. Hereafter, this peak is called
the `second peak'. This peak is achieved for materials in the range of
density $10^7$ -- $10^{7.5}$~$\gcm$ in all the models. At the second
peak, model \mdlx{1.4} achieves the highest smoothed temperature. The
raw and smoothed temperature at the second peak converge to $2.2
\times 10^9$~K and $1.5 \times 10^9$~K, respectively.

Table~\ref{tab:maximum_temperature} summarizes the properties of the
first and second peaks. The first peaks of the raw and smoothed
temperatures are not coincident. Since the time lag between the first
peaks is at most $5$~s, being sufficiently smaller than the orbital
time ($22$~s), these peaks appear at the same merging phases. This is
true for the second peaks of the raw and smoothed temperatures, except
for model \mdlx{11}. However, in model \mdlx{11}, the smoothed
temperature at the time of the second peak of the raw temperature gets
as high as $\sim 1.4 \times 10^9$~K, comparable to the smoothed
temperature at its second peak. Accidentally, the former smoothed
temperature is slightly smaller than the latter smoothed temperature.

\begin{table}
  \caption{Properties of the first and second peaks: their type of
    temperature, temperature, density, and
    time.} \label{tab:maximum_temperature}
  \begin{center}
    \begin{tabular}{r|ccccc}
      Model & Peak & Type & $T$ [$10^9$~K] & $\rho$ [$10^6$~gcm$^{-3}$] & Time [s] \\
      \hline
      \hline
      \mdlx{1.4} & 1st & $T_{{\rm r},i}$ & 2.5 & 1.7 & 284 \\
                 & 1st & $T_{{\rm s},i}$ & 1.3 & 6.4 & 279 \\
                 & 2nd & $T_{{\rm r},i}$ & 1.9 & 13  & 344 \\
                 & 2nd & $T_{{\rm s},i}$ & 1.4 & 6.4 & 346 \\
      \hline
      \mdlx{2.8} & 1st & $T_{{\rm r},i}$ & 3.4 & 2.9 & 194 \\
                 & 1st & $T_{{\rm s},i}$ & 1.5 & 3.6 & 195 \\
                 & 2nd & $T_{{\rm r},i}$ & 2.1 & 13  & 230 \\
                 & 2nd & $T_{{\rm s},i}$ & 1.3 & 5.0 & 226 \\
      \hline
      \mdlx{5.5} & 1st & $T_{{\rm r},i}$ & 3.8 & 2.9 & 134 \\
                 & 1st & $T_{{\rm s},i}$ & 1.6 & 1.9 & 133 \\
                 & 2nd & $T_{{\rm r},i}$ & 2.1 & 19  & 171 \\
                 & 2nd & $T_{{\rm s},i}$ & 1.5 & 9.2 & 173 \\
      \hline
      \mdlx{11}  & 1st & $T_{{\rm r},i}$ & 3.8 & 3.8 & 110 \\
                 & 1st & $T_{{\rm s},i}$ & 2.1 & 3.7 & 111 \\
                 & 2nd & $T_{{\rm r},i}$ & 2.2 & 13  & 138 \\
                 & 2nd & $T_{{\rm s},i}$ & 1.4 & 10  & 153 \\
    \end{tabular}
  \end{center}
\end{table}

At $t=500$~s, the raw and smoothed temperatures converge to $9 \times
10^8$~K and $8 \times 10^8$~K, respectively, for materials whose
density is in the range of $10^{5}$ -- $10^{7}$~$\gcm$. As for the
range of density $10^{7}$ -- $10^{7.5}$~$\gcm$, the smoothed
temperature converges to $6 \times 10^8$~K, while the raw temperature
does not. This is because the raw temperature in model \mdlx{5.5}
jumps up at $t=350$~s. This jump-up is due to an effect of an
artificial viscosity adopted in the SPH simulations. Even a slight
amount of the artificial viscosity can highly increase temperature of
a particle in a high density region, since temperature is sensitive to
an internal energy especially in the high density region. This effect
is not seen in models \mdlx{1.4} and \mdlx{2.8}, and therefore the raw
temperature in these models rather than in model \mdlx{5.5} should be
correct at $t=500$~s. Therefore, the raw temperature is $8 \times
10^8$~K at the density of $10^{7}$ -- $10^{7.5}$~$\gcm$ at
$t=500$~s. When we compare temperatures of materials at the density
$<10^7$~$\gcm$ and $>10^7$~$\gcm$, the former temperature is higher
than the latter temperature. This is because a lower density region is
shock-heated more strongly during the merger event.

\begin{figure*}
  \begin{center}
    \includegraphics[width=12.0cm]{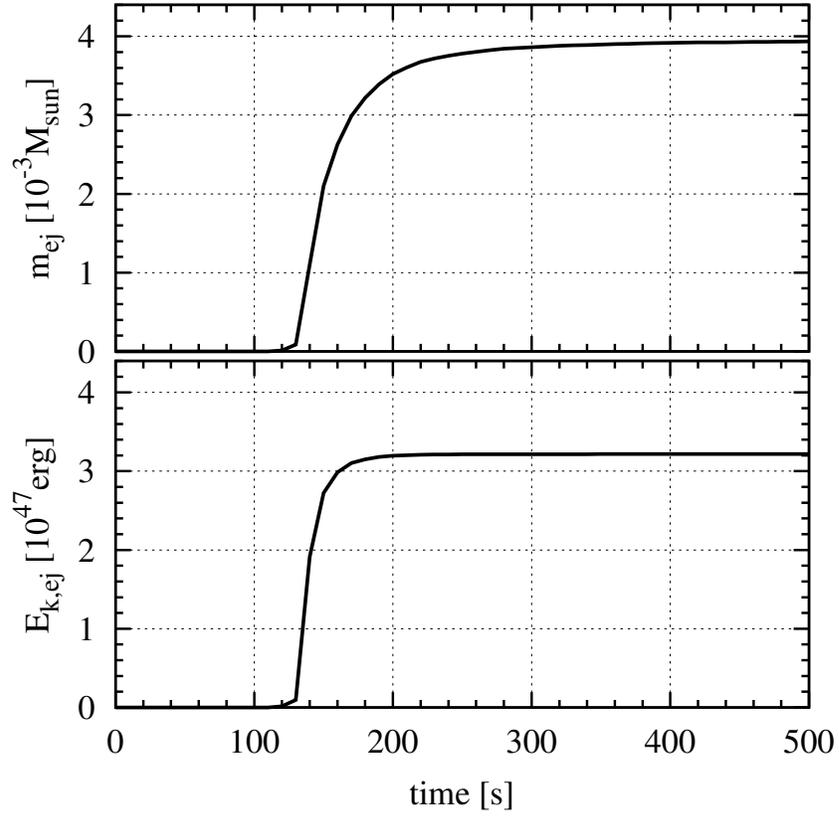}
  \end{center}
  \caption{Time evolution of mass (top) and kinetic energy (bottom) of
    merger ejecta in model \mdlx{5.5}.}
  \label{fig:ejecta_time}
\end{figure*}

Below, we discuss the properties of merger ejecta in model
\mdlx{5.5}. Figure~\ref{fig:ejecta_time} shows the time evolution of
the mass and kinetic energy of the merger ejecta. The mass is small at
the merger time ($t=120$~s), being only $1.3 \times
10^{-5}M_{\odot}$. The ejecta mass then increases rapidly during
$t=140$ -- $150$~s. Finally, the ejecta mass reaches $3.9 \times
10^{-3} M_{\odot}$ at $t=500$~s. At $t=500$~s, the ejecta stops
growing in mass. The ejecta mass reaches a constant value at about
$t=500$~s, while the kinetic energy does so at an earlier time, about
$t=200$~s. This is because the merger ejecta with higher velocities
are formed and ejected at an earlier time. Eventually, their total
kinetic energy at infinity reaches $\sim 3.2 \times 10^{47}$~erg.

Table~\ref{tab:ejecta} shows the tidal mass, and the masses and
kinetic energies of the total ejecta in models \mdlx{1.4}, \mdlx{2.8},
and \mdlx{5.5}. We can not investigate these properties of the total
ejecta in model \mdlx{11}, since we do not follow the evolution of
this model until a dynamical steady state is reached. The properties
of the ejecta are independent of the mass resolution.

\begin{table*}
  \caption{The tidal mass, and the mass and kinetic energy of the
    total ejecta in each model.} \label{tab:ejecta}
  \begin{center}
    \begin{tabular}{r|ccc}
      Model & tidal ejecta mass & Total mass [$M_\odot$] & Total kinetic energy [erg] \\
      \hline
      $1.4$M  & $2.1 \times 10^{-5}$ & $5.0 \times 10^{-3}$ & $3.0 \times 10^{47}$ \\
      $2.8$M  & $2.1 \times 10^{-5}$ & $3.9 \times 10^{-3}$ & $3.1 \times 10^{47}$ \\
      $5.5$M  & $1.3 \times 10^{-5} $& $3.9 \times 10^{-3}$ & $3.2 \times 10^{47}$ \\
      $11$M   & --                   & --                   & --
    \end{tabular}
  \end{center}
\end{table*}

\begin{figure*}
  \begin{center}
    \includegraphics[width=12.0cm]{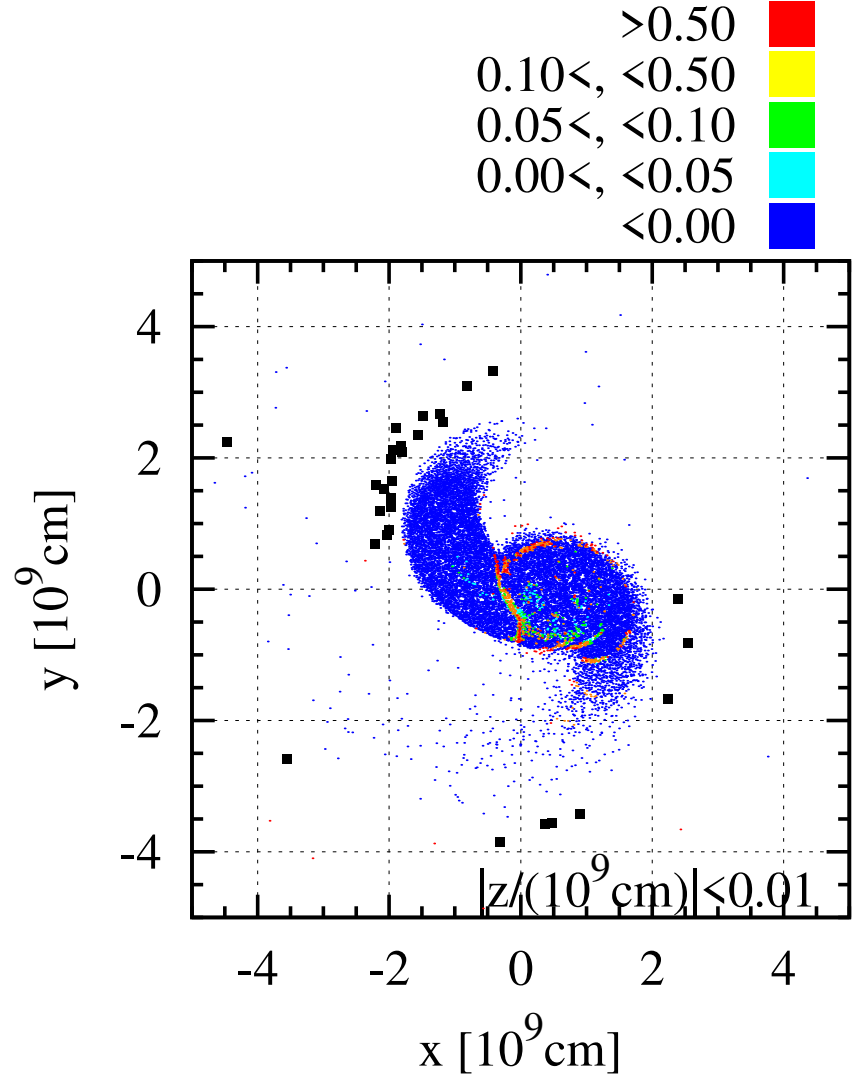}
  \end{center}
  \caption{Shock detector distribution at $t=120$~s. Particles with
    $|z|<10^7$~cm are drawn. Black dots indicate merger ejecta at
    $t=120$~s.}
  \label{fig:ejecta_tide}
\end{figure*}

The merger ejecta can be divided into two groups, depending on their
formation mechanism. The merger ejecta in the first group can be seen
in Figure~\ref{fig:ejecta_tide}. They are generated from tidal tails
of the primary and companion, and so they are called `tidal
ejecta'. They become unbound, since they receive orbital angular
momenta from rotating bar potential formed by the binary system. They
are formed only before $t=130$~s. The total mass of them is $8.6
\times 10^{-5}M_{\odot}$. The tidal ejecta become unbound not due to a
shock wave, since they are far away from shocked regions. The shocked
region is identical to that creating hotspots at Pakmor's time (see
section~\ref{sec:violent}). They are not strong shock, since their
shock detectors are a bit smaller than unity.

\begin{figure*}
  \begin{center}
    \includegraphics[width=16.0cm]{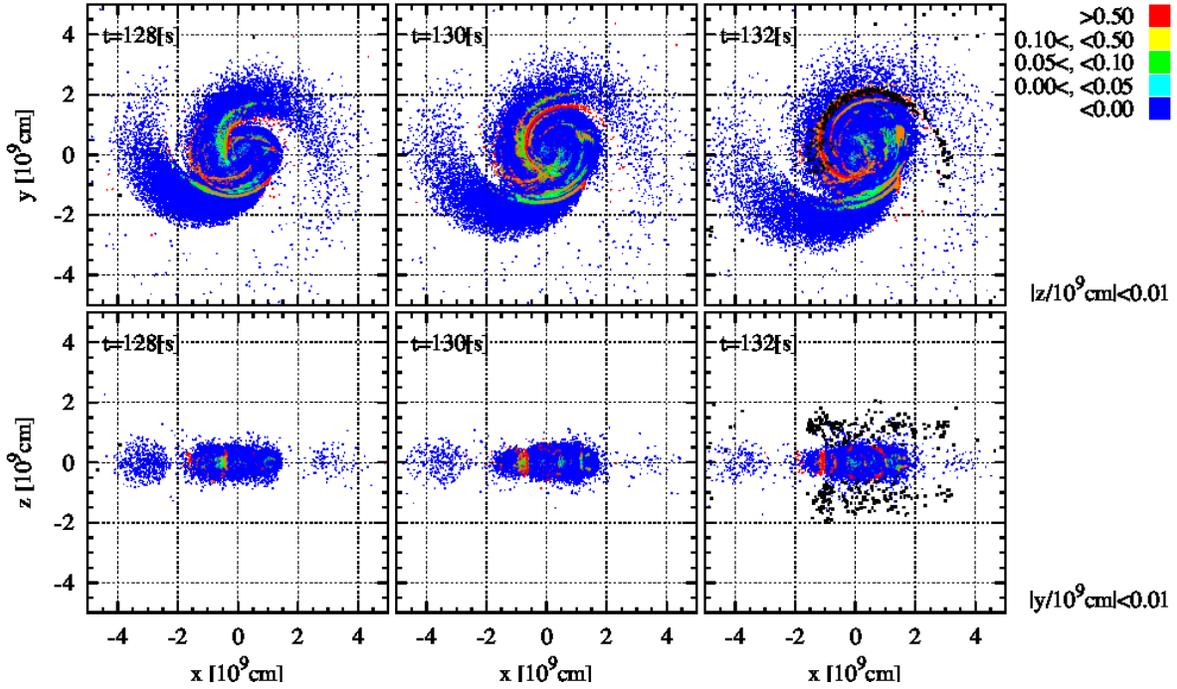}
  \end{center}
  \caption{Shock detector distributions at $t=128$, $130$, and
    $132$~s. Particles with $|z|<10^7$~cm and with $|y|<10^7$~cm are
    drawn in the top and bottom panels, respectively. Black dots
    indicate merger ejecta at $t=130$~s.}
  \label{fig:ejecta_shck}
\end{figure*}

In Figure~\ref{fig:ejecta_shck}, we show the moment when a part of the
merger ejecta in the second group are formed. They are generated in a
shocked region (see the top panels). Hereafter, they are called
`shocked ejecta'. The shocked region arises from a collision between
the main body of the system and a tidal tail. The merger ejecta are
prevented from traveling toward the directions of the orbital plane,
which can be seen in the bottom right panel. This is because the tidal
tail acts as an obstacle. The shocked ejecta are formed several times
through the above mechanism after $t=130$~s. They dominate the total
mass of the merger ejecta.

The terminal velocities of the tidal ejecta are typically $3$ -- $4
\times 10^8~\cms$, similar to the relative velocity between the
primary and companion just before their merger. Those of the shocked
ejecta range from $10^7~\cms$ to $10^9~\cms$. These velocities are the
largest just after the merger, and decrease gradually.

\subsection{Comparison with previous studies}
\label{sec:comparison}

In this section, we compare our results with those obtained by
previous studies. First, we use the results of a binary with masses of
$1.1M_{\odot}$ and $1.0M_{\odot}$, and next use those of a binary with
masses of $0.9M_{\odot}$ and $0.6M_{\odot}$.

For a binary with masses of $1.1M_{\odot}$ and $1.0M_{\odot}$, we
focus on the raw temperature and the mass of the merger ejecta at its
final state. We also check the smoothed temperature, if it is
available for comparison in any of the previous studies.  Not all the
previous studies treat binary CO~WDs exactly with masses of $1.1$ and
$1.0M_{\odot}$, and for such cases we adopt, for comparison to work
out, previous studies with mass combinations similar to the one in our
simulations.  In our results, the maximum raw and smoothed
temperatures are, respectively, $9 \times 10^8$~K and $8 \times
10^8$~K (see section~\ref{sec:overview}). The corresponding
temperature in previous studies is as follows: In \cite{Dan14}, the
maximum raw and smoothed temperatures are, respectively, $11 \times
10^8$~K and $8.5 \times 10^8$~K in the case of binary CO~WDs with
masses of $1.05$ and $1M_{\odot}$ (see their table~A1). In
\cite{Zhu13}, their maximum raw temperature is $9.4 \times 10^8$~K in
binary CO~WDs with masses of $1.0$ and $1.0M_{\odot}$ (see the column
of $(T_{\rm max}^z)_8$ in their table~2). In \cite{Raskin12}, the
maximum raw temperature is about $12.5 \times 10^8$~K for a binary
with masses of $1.06$ and $0.96M_{\odot}$ (see their figure~9). Our
results are in good agreement with theirs.

The total mass of the merger ejecta in our simulation is $3.9 \times
10^{-3} M_{\odot}$. In \cite{Dan14}, it is $7.64 \times 10^{-4}
M_{\odot}$ for a binary with masses of $1.0M_{\odot}$ and
$1.0M_{\odot}$, $8.50 \times 10^{-4} M_{\odot}$ for $1.05M_{\odot}$
and $1.0M_{\odot}$, and $1.453 \times 10^{-3} M_{\odot}$ for
$1.05M_{\odot}$ and $1.05M_{\odot}$. Our mass is consistent with those
in \cite{Dan14} to the first order.

In our results, the total mass of the tidal ejecta is $8.6 \times
10^{-5}M_{\odot}$. In \cite{Raskin13}, the mass of the tidal ejecta is
$4.7 \times 10^{-3} M_{\odot}$ for a binary with masses of
$1.06M_{\odot}$ and $1.06M_{\odot}$, and $3.3 \times 10^{-3}
M_{\odot}$ for $1.20M_{\odot}$ and $1.06M_{\odot}$ (see their
table~1). According to a fitting formula of \cite{Dan14} (see their
eq. A11), the mass of the tidal ejecta is $\sim 1 \times 10^{-3}
M_\odot$ in the case of a binary with masses of $1.1M_{\odot}$ and
$1.0M_{\odot}$.  Our mass is smaller than those in these previous
studies by an order of magnitude. The small merger time of model
\mdlx{5.5} should not affect the small mass of the tidal ejecta. In
model \mdlx{1.4}, the CO~WDs orbit around each other more than ten
times, but the mass of the tidal ejecta is less than $2.1 \times
10^{-5}M_{\odot}$, which is almost the same as in model \mdlx{5.5}
(see Table~\ref{tab:ejecta}). Since the tidal ejecta are a minor
component, the formation of them would be sensitive to the detail of
the setup of simulations. We do not discuss this discrepancy anymore,
but we should keep in mind that we may underestimate the total mass of
the tidal ejecta.

\begin{figure*}
  \begin{center}
    \includegraphics[width=12.0cm]{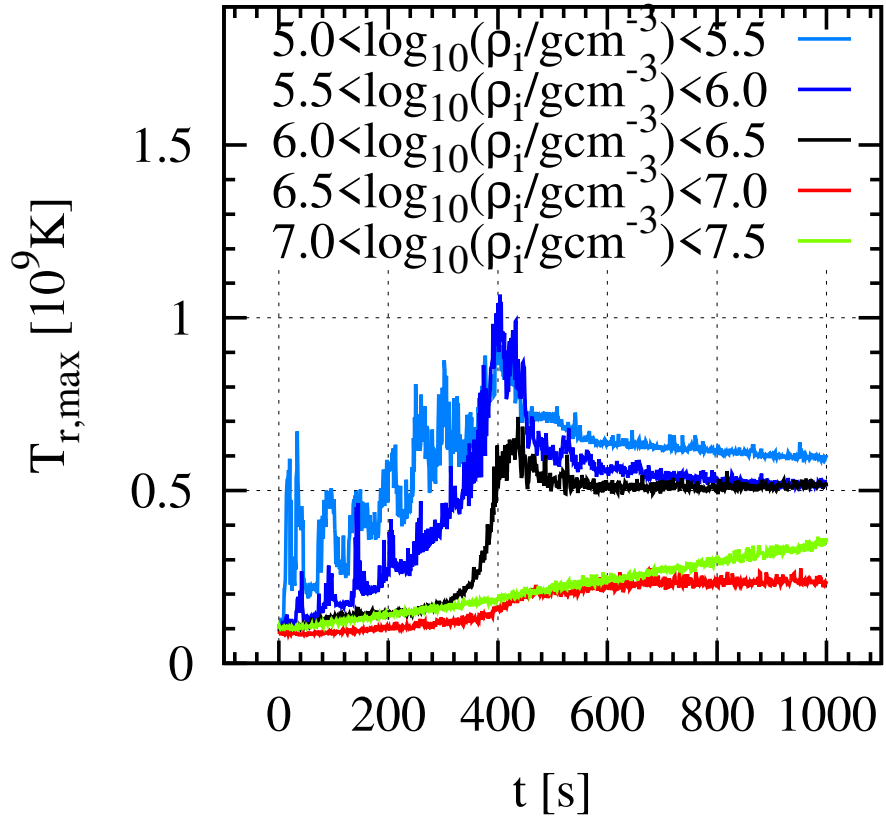}
  \end{center}
  \caption{Time evolution of maximum raw temperature at ranges of mass
    density in the case of binary CO~WDs with $0.9$ and
    $0.6M_{\odot}$. The ranges of mass density are indicated in the
    panel.}
  \label{fig:bench_ov}
\end{figure*}

For a binary with masses of $0.9M_{\odot}$ and $0.6M_{\odot}$, we show
the time evolution of the raw temperature in
Figure~\ref{fig:bench_ov}. The maximum raw temperature after the
merger event ($t=1000$~s) is $6 \times 10^8$~K or less depending on
the position of the materials within the final merger remnant. This is
consistent with those found in previous studies
\citep{Yoon07,Zhu13,Dan14}. In \cite{Nomoto13}, the corresponding
temperature was reported to be $8.5 \times 10^8$~K. This is a bit
higher than our results although \cite{Nomoto13} adopted the same
simulation code as ours. We find that this difference comes from the
difference in the recipes making the initial conditions. We relax a
configuration of the binary CO~WDs, while \cite{Nomoto13} did
not. This is consistent with the argument by \cite{Dan11}, in which
the maximum raw temperature at the final state becomes higher without
the relaxation.

For this set of the binary parameters ($0.9M_{\odot}$ and
$0.6M_{\odot}$), the total mass in the merger ejecta is about $8.7
\times 10^{-4}M_{\odot}$, which consists mainly of tidal ejecta. The
ejecta mass of a binary with masses of $0.9M_{\odot}$ and
$0.6M_{\odot}$ is much smaller than that of a binary with masses of
$1.1M_{\odot}$ and $1.0M_{\odot}$, since the former merger is less
violent than the latter. The corresponding masses are $2.0 \times
10^{-3}M_{\odot}$ in binary CO~WDs with masses of $0.96$ and
$0.64M_{\odot}$ in \cite{Raskin13}, and $1.0 \times 10^{-3}M_{\odot}$
for a combination of $0.9$ and $0.65M_{\odot}$ in \cite{Dan14} from
their table~A1. Our result is almost consistent with the previous
studies.

In summary, our raw temperature in the final state is consistent with
those of the previous studies. The total mass of the merger ejecta is
also in good agreement with the previous studies. However, the mass of
the tidal ejecta is different from those found in previous studies by
an order of magnitude.  Since the tidal ejecta are a minor component,
the formation of them would be sensitive to the detail of the setup of
simulations.

\section{Assessment of explosion models}
\label{sec:verification}

In this section, we assess explosion models based on the results of
our SPH simulations. These models are the helium-ignited violent
merger, carbon-ignited violent merger, Chandrasekhar mass (see
section~\ref{sec:intro}). We also discuss other branches of models
which are categorized in none of these models.  We investigate these
explosion models in chronological order: The helium-ignited violent
merger model (section~\ref{sec:double}), the carbon-ignited violent
merger model (section~\ref{sec:violent}), the other models
(section~\ref{sec:other}), and the Chandrasekhar mass model
(section~\ref{sec:chandrasekhar}). In section~\ref{sec:double},
\ref{sec:violent}, and \ref{sec:chandrasekhar}, we investigate whether
binary CO~WDs can lead to an explosion, and whether the explosion can
be observed as an SN~Ia. In section~\ref{sec:other}, we only discuss
what an explosion should look like, assuming the explosion is
triggered by these models.

\subsection{Helium-ignited violent merger model}
\label{sec:double}

In this model, the helium detonation at the surface of the primary
induces compression of the core materials of the primary. This
triggers the carbon detonation in the core, subsequently leading to an
explosion. In order to assess a possibility of such a mode in the
explosion, we provide two check points; whether the helium detonation
is initiated, and whether it triggers the carbon detonation.

We define conditions of initiation of the helium detonation. The
helium detonation is a supersonic flame powered by helium burning,
i.e. the triple-alpha reaction. In order for the triple-alpha reaction
to power the flame, the triple-alpha reaction needs to proceed faster
than cooling due to an adiabatic expansion. In other words, the
timescale of the triple-alpha reaction should be shorter than the
local dynamical timescale of the fluid under consideration.
Therefore, we adopt $\tta < \tdy$ as a criterion for the initiation
condition of the helium detonation. We should keep in mind that this
is only a necessary condition. For a necessary and sufficient
condition, a condition that the flame propagates supersonically must
be satisfied.  However, we do not consider such a condition in this
paper.

\begin{figure*}
  \begin{center}
    \includegraphics[width=12.0cm]{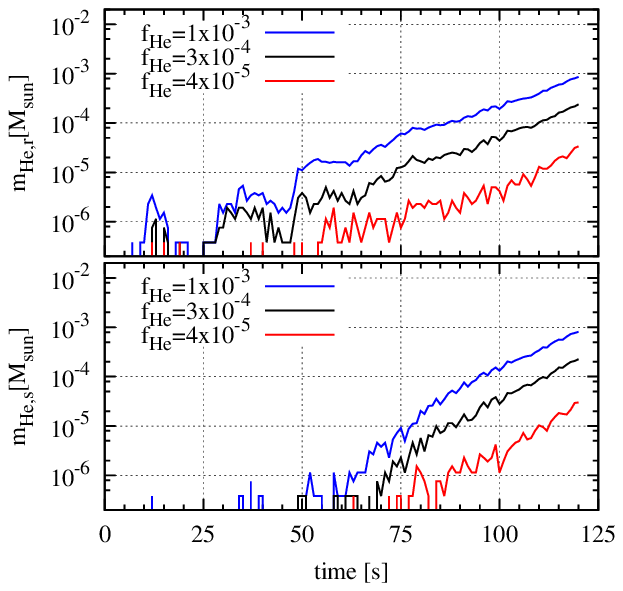}
  \end{center}
  \caption{Time evolution of the total mass of helium particles with
    $\ttar < \tdy$ (top) and $\ttas < \tdy$ (bottom).}
  \label{fig:helium_mass}
\end{figure*}

Figure~\ref{fig:helium_mass} shows the time evolution of the total
mass of helium particles with $\tta < \tdy$ in model \mdlx{5.5}. The
total mass becomes non-zero before the merger time ($t=120$~s),
regardless of a choice of $\ttar$ or $\ttas$, and regardless of the
value of $\fhe$. Therefore, the system generally satisfies the
necessary condition to initiate the helium detonation before the
merger time.

\begin{figure*}
  \begin{center}
    \includegraphics[width=12.0cm]{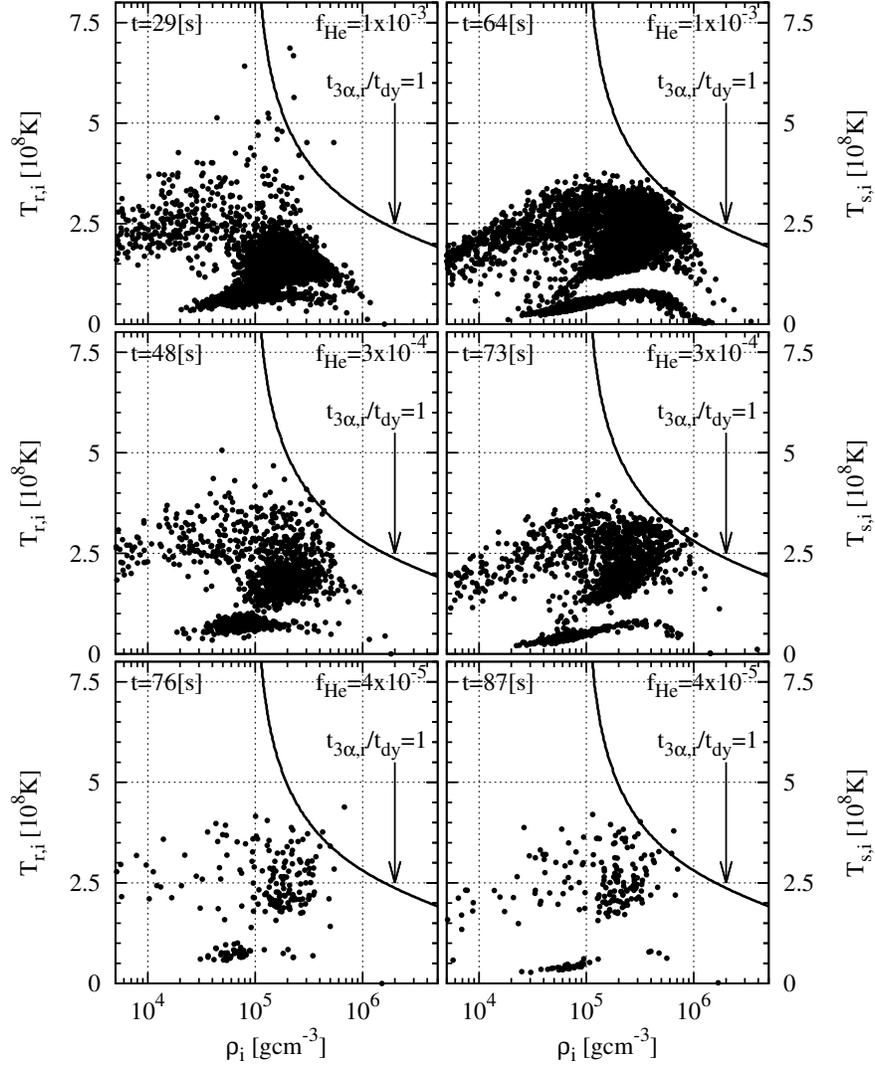}
  \end{center}
  \caption{Mass density and temperature of helium particles at the
    time when the helium detonation is initiated. In the left and
    right panels, vertical axes indicate the raw and smoothed
    temperatures, respectively. From top to bottom, $\fhe = 1 \times
    10^{-3}$, $3 \times 10^{-4}$, and $4 \times 10^{-5}$. A curve in
    each panel shows mass density and temperature at which timescale
    of the triple-alpha reaction is equal to local dynamical
    timescale.}
  \label{fig:helium_rhot}
\end{figure*}

We define the initiation time of the helium detonation as the time
after which the total mass of helium particle with $\tta < \tdy$
raises more than $10^{-6}M_\odot$ and keeps to be so, in order to
exclude numerical noises. The initiation time in all the cases is
summarized in Table~\ref{tab:helium_time}. The initiation time is
earlier when we adopt $\ttar$ and a larger value of $\fhe$. We can
also see Figure~\ref{fig:helium_rhot} that a few helium particles have
$\tta < \tdy$ at the initiation times of the helium detonation in all
the cases.

We ignore the contribution of the alpha process, such as
$^{12}$C($\alpha, \gamma$)$^{16}$O, to the initiation of the helium
detonation. This is because the triple-alpha reaction is much faster
than the alpha process when temperature is less than $10^9$~K
\citep[e.g.][]{Shen14}. The alpha process is important for the
propagation of the helium detonation, but unimportant for the
initiation of the helium detonation.

The small merger time in our simulation may artificially increase the
temperature of the helium particles, since the merging process may
start suddenly. We thus note that the helium detonation might easily
be initiated artificially. However, we do not consider this effect in
the following discussion.

\begin{table}
  \caption{The initiation time of the helium detonation, and expected
    \fsni mass as a result of the detonation.} \label{tab:helium_time}
  \begin{center}
    \begin{tabular}{cccc}
      \hline
      temperature & $\fhe$ & time [s] & \fsni [$M_\odot$] \\
      \hline
      \hline
      raw & $1 \times 10^{-3}$ & $29$ & $1.1 \times 10^{-6}$ \\
      raw & $3 \times 10^{-4}$ & $48$ & $1.1 \times 10^{-6}$ \\
      raw & $4 \times 10^{-5}$ & $76$ & $3.8 \times 10^{-7}$ \\
      \hline
      smoothed & $1 \times 10^{-3}$ & $64$ & $1.1 \times 10^{-5}$ \\
      smoothed & $3 \times 10^{-4}$ & $73$ & $3.4 \times 10^{-6}$ \\
      smoothed & $4 \times 10^{-5}$ & $87$ & $3.8 \times 10^{-7}$ \\
      \hline
    \end{tabular}
  \end{center}
\end{table}

Next, we consider whether the helium detonation can trigger the carbon
detonation in the core of the primary. The helium detonation is
thought to trigger the carbon detonation in either of two ways. In one
way, the helium detonation hits the core of the primary, and directly
ignites the carbon detonation there.  This is called a `direct drive'
mechanism. In the other way, the shock wave created by the helium
detonation propagates into the core with little or no carbon burning.
If the helium detonation region encloses the core, the shock wave then
could converge somewhere in the core, and could become sufficiently
strong to initiate the carbon detonation.  This is called a
`converging shock' mechanism. For the direct drive mechanism, a
substantial amount of helium is required \citep[e.g.][]{Moll13}. It is
unlikely that this mechanism triggers the carbon detonation in our
models, since $\fhe$ is small. We therefore focus on the converging
shock mechanism, to assess if our system could lead to an explosion by
this mode, i.e., the helium-ignited violent merger.

\begin{figure*}
  \begin{center}
    \includegraphics[width=12.0cm]{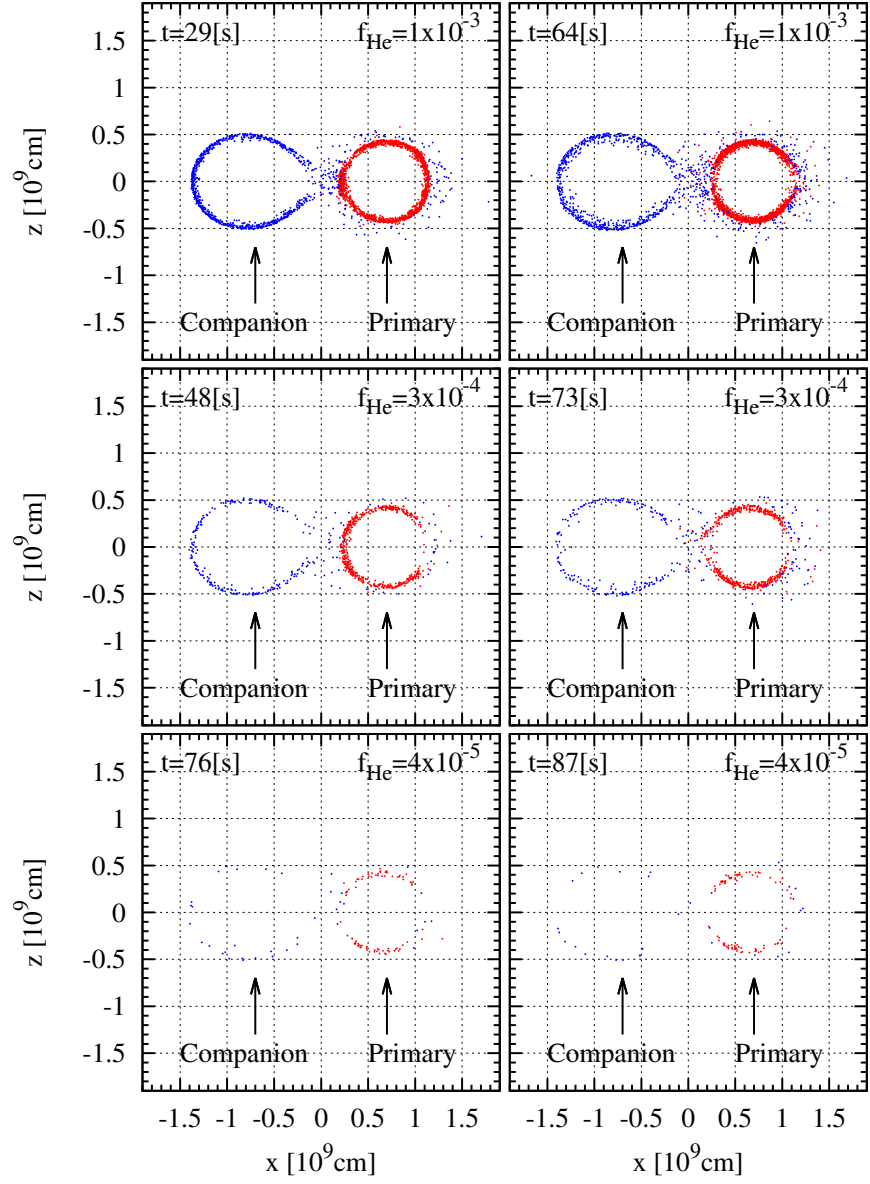}
  \end{center}
  \caption{Distribution of helium particles with $|y| < 2 \times
    10^8$~cm at the initiation times of the helium detonation. Red and
    blue points indicate helium particles originating from the primary
    and companion, respectively. From top to bottom, $\fhe = 1 \times
    10^{-3}, 3 \times 10^{-4}$ and $4 \times 10^{-5}$. In the left and
    right panels, the raw and smoothed temperatures are chosen for the
    initiation conditions of the helium detonation, respectively.}
  \label{fig:helium_view}
\end{figure*}

For the converging shock mechanism to successfully operate, a helium
layer needs to enclose the core of the primary. We investigate the
distribution of helium at the initiation time of the helium
detonation, as shown in Figure~\ref{fig:helium_view}.  In the cases of
$\fhe = 1 \times 10^{-3}$ and $3 \times 10^{-4}$, the helium particles
enclose the core of the primary, regardless of a choice of the raw or
smoothed temperature to define the initiation time of the helium
detonation.  However, in the cases of $\fhe = 4 \times 10^{-5}$, the
helium particles are sparse on the orbital plane, again regardless of
a choice of the raw or smoothed temperature.

A reason for this sparseness can be explained as follows. The mass
accretion from the companion to the primary proceeds in the following
way; the helium particles first fall, and subsequently the
carbon-oxygen particles do so. The carbon-oxygen particles have
already started falling onto the surface of the primary at the
initiation time of the helium detonation. These carbon-oxygen
particles hitting the surface of the primary on the orbital plane push
the helium particles away. When $\fhe$ is smaller, the helium
particles are distributed more sparsely on the orbital plane at the
initiation time of the helium detonation for two reasons: First,
because of the smaller $\fhe$, all the helium particles are more
easily pushed away from the orbital plane. Second, as $\fhe$ becomes
smaller, the initiation time becomes closer to the merger time (see
Table~\ref{tab:helium_time}). As the initiation time is closer to the
merger time, the carbon-oxygen particles are accreted by the primary
more violently (see Figure~\ref{fig:hotspot_ov}), pushing the helium
particles away more easily.

From the above, we conclude that only in the case of $\fhe \gtrsim 3
\times 10^{-4}$, the helium detonation can potentially succeed in
triggering the core carbon detonation, and can lead to an
explosion. In \cite{Pakmor13}, the helium detonation encloses the core
of their primary in $\fhe = 0.01$. This result is consistent with our
results.

\begin{figure*}
  \begin{center}
    \includegraphics[width=10.0cm]{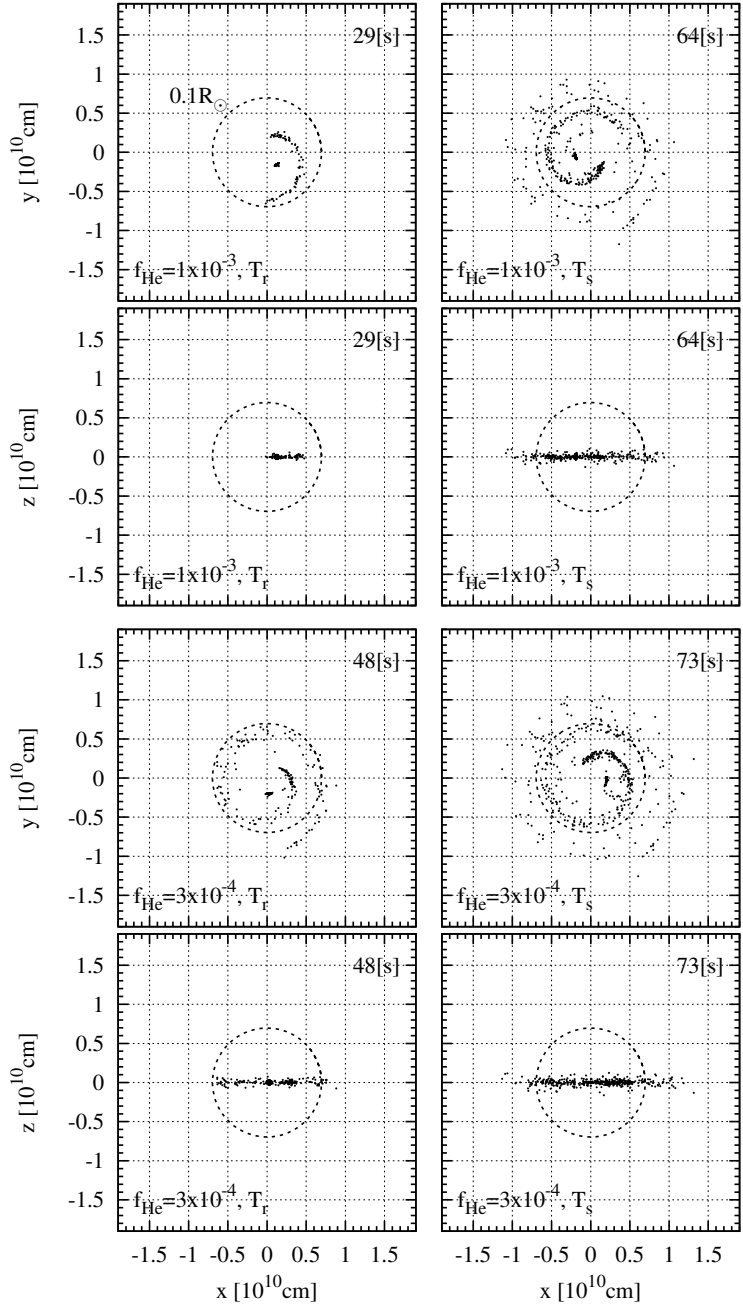}
  \end{center}
  \caption{Material distribution of a binary with masses of
    $1.1M_\odot$ and $1.0M_\odot$ in model \mdlx{5.5}. We draw only
    particles separated from the center of the primary by $2 \times
    10^9$~cm. The time is indicated at the top right in each
    panel. The time is the initiation time of the helium detonation in
    the cases of $\fhe = 1 \times 10^{-3}$ and the raw temperature
    (two top-left panels), $\fhe = 1 \times 10^{-3}$ and the smoothed
    temperature (two top-right panels), $\fhe = 3 \times 10^{-4}$ and
    the raw temperature (two bottom-left panels), and $\fhe = 3 \times
    10^{-4}$ and the smoothed temperature (two bottom-right
    panels). Dashed curves indicate $0.1R_\odot$ from the center of
    the primary.}
  \label{fig:debris_sdis_helium}
\end{figure*}

As described above, our binary can potentially explode through the
helium-ignited violent merger model, if $\fhe \gtrsim 3 \times
10^{-4}$. Hereafter we consider expected observational outcome,
focusing on brightness just after the explosion, hereafter called
``early brightness''.  In this phase, the optical photons are
basically powered by the thermal energy content produced by the shock
heating following the SN explosion, in the envelope of the exploding
star.  This energy reservoir is sensitive to the structure of the
binary, especially to the size of the envelope of the exploding
progenitor.  Figure~\ref{fig:debris_sdis_helium} shows the matter
distribution of our binary system at the initiation time of the helium
detonation. It has a tidal-tail structure at $t=29$ and $48$~s, and a
disk structure at $t=64$ and $73$~s. The materials spread out to $\sim
0.1R_{\odot}$ away from the center of the primary.

The early brightness is also affected by a radius of the companion,
through the interaction between the SN ejecta and the companion. The
radius of the companion at the initiation time of the helium
detonation is similar to that at the initial time. Then, the companion
radius is about $5 \times 10^8$~cm ($7 \times 10^{-3}R_\odot$).

Moreover, the early brightness can be partly powered by radioactive
decay of $^{56}$Ni synthesized by the helium detonation. We estimate
an amount of $^{56}$Ni as follows. The helium detonation synthesizes
$^{56}$Ni when the density is higher than a critical density, $\sim
10^7~\gcm$, after the helium detonation passes. Taking into account
the shock compression, we assume that the helium particles with
$\rho_i>10^6~\gcm$ at the initiation time of the helium detonation are
converted to $^{56}$Ni. We show the amount of $^{56}$Ni as estimated
in this way, in all the cases in the fourth column of
Table~\ref{tab:helium_time}.

We qualitatively compare the expected early brightness resulting from
the helium-ignited violent merger model of our binary with the early
brightness of SN~2011fe and SN~2014J, in terms of radii of the primary
and companion at the time of the helium ignition, and \fsni mass after
the helium detonation. The expected light curve will be much fainter
than the light curve of SN~2014J. Following the analysis of its early
brightness \citep{Goobar14}, SN~2014J is suggested to have either a
large primary radius ($\sim 1R_\odot$), a larger companion radius
($\sim 4R_\odot$), or a large amount of \fsni mass near the surface
($10^{-3}M_{\odot}$). All of these are larger than found in the
results of our simulation. Another test is provided by SN~2011fe.  The
expected early brightness may likely be brighter than the light curve
of SN~2011fe; the progenitor of SN~2011fe is suggested to have a small
radius, $< 0.1R_\odot$\citep{Nugent11,Bloom12,Zheng13,Mazzali14}.
However, we should keep in mind that our `envelope' has tidal-tail and
disk-like structures, which has not been taken into account in models
to connect the size of the progenitor and the early brightness. This
structure should affect the expected early brightness; the light curve
may be fainter than we expect above because of the small opening angle
of the envelope from the primary, and may be consistent with that of
SN~2011fe. In order to quantitatively compare the early brightness
resulting from our binary with those of SNe 2014J and 2011fe, we have
to follow the explosion of our binary by means of numerical
simulations. This is our future work.

Another test is provided by properties of SN remnants.
\cite{Papish14} have investigated expected properties of SN ejecta
from an exploding CO~WD with a helium WD companion. In their model,
the separation is $0.08R_\odot$, and the helium WD has a radius of
$0.02R_\odot$. The resulting SN remnant does not have a spherically
symmetric shape, since the ejecta of the SN prevents from moving
beyond the helium WD in this direction.  We anticipate that an SN
remnant resulting from the explosion of our binary through the
helium-ignited violent merger model has a similar shape to that of
\cite{Papish14}. This is because their and our binaries have similar
opening angles of the companion from the primary, i.e., $14$ degree in
\cite{Papish14} and $19$ degree in our case. The SN remnant with a
non-spherically symmetric shape is generally not consistent with a
large fraction of SN~Ia remnants, which tend to be spherically
symmetric.

This section is summarized as follows: When $\fhe \gtrsim 3 \times
10^{-4}$, our binary can explode through the helium-ignited violent
merger model. However, the expected explosion has different features
from SN~Ia in several respects.  The explosion has the early
brightness much brighter than that of SN~2011fe, and much fainter than
that of SN~2014J. Note that the light curve may be fainter than we
expect, and may be consistent with that of SN~2011fe. In addition, the
explosion will lead to an SN remnant with a non-spherically symmetric
shape, which is inconsistent with shapes of a large fraction of SN~Ia
remnants.  We note that it is unclear whether the other mass
combinations can explode through this mode, and whether the explosions
can be observed as an SN~Ia.

\subsection{Carbon-ignited violent merger model}
\label{sec:violent}

If the system survives with no or insufficient energy injection from
the helium detonation, there is a chance that the carbon detonation
initiated at a hotspot leads to an explosion, i.e., the carbon-ignited
violent merger model. A successful explosion in this mode depends on
whether the hotspot appears in which the nuclear reaction proceeds
rapidly to lead to the carbon detonation. A necessary condition for
this mode to lead to a successful explosion is the following: For the
carbon detonation to take place, the $^{12}$C + $^{12}$C reaction
should proceed rapidly to lead to thermonuclear runaway. Therefore, a
heating rate by the $^{12}$C + $^{12}$C reaction should exceed a
cooling rate due to an adiabatic expansion. In other words, timescale
of the $^{12}$C + $^{12}$C reaction should be shorter than local
dynamical timescale, i.e. $\tcc < \tdy$. We call materials which
satisfy this condition the hotspots.

\begin{figure*}
  \begin{center}
    \includegraphics[width=16.0cm]{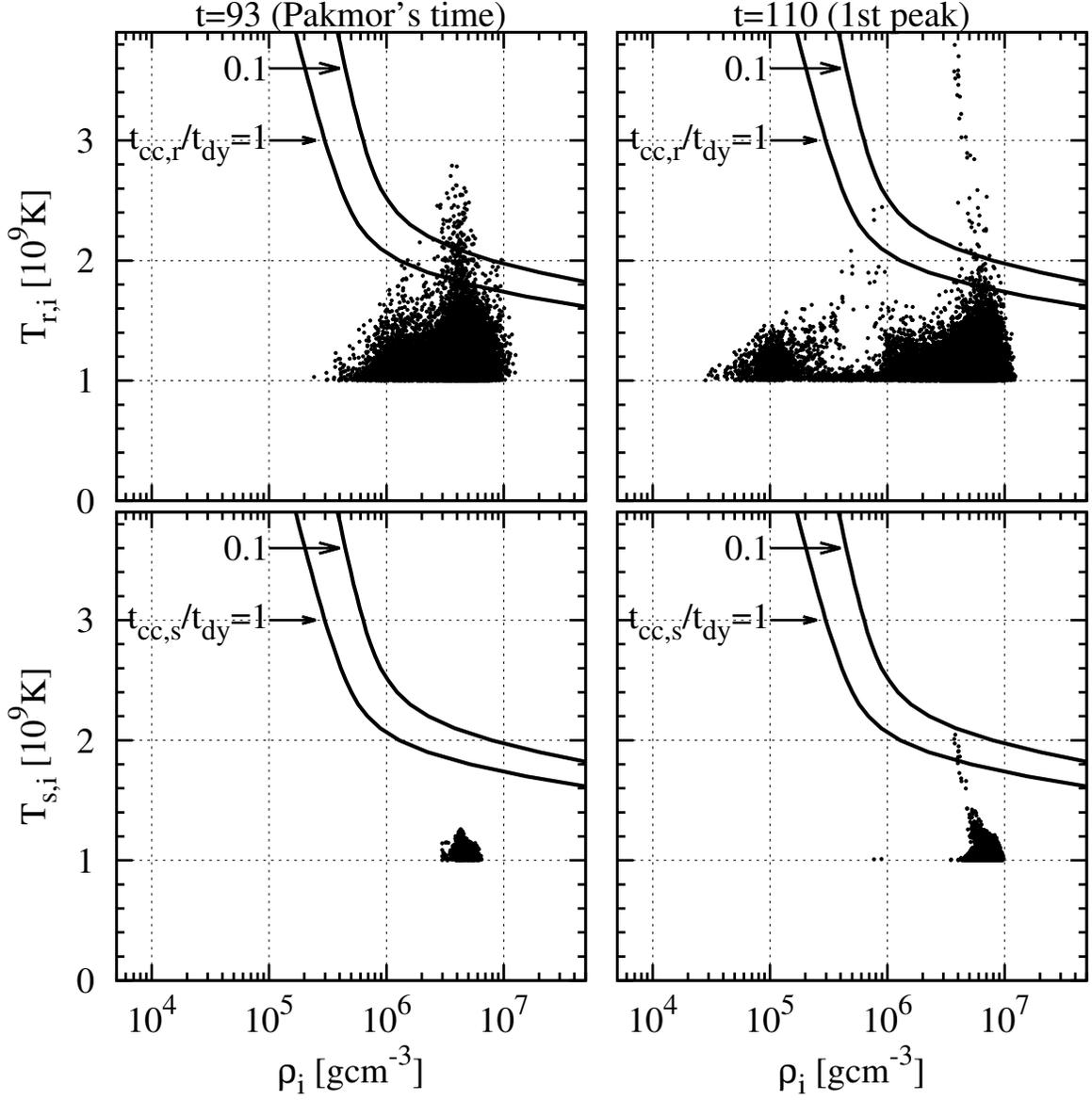}
  \end{center}
  \caption{Mass density and temperature of each particle in model
    \mdlx{11} at $t=93$ (left) and $t=110$ (right), which are
    corresponding to Pakmor's time and the time at the first peak. In
    the top and bottom panels, the vertical axes show raw and smoothed
    temperatures, respectively. Particles with $T_{{\rm r},i}<10^9$~K
    and with $T_{{\rm s},i}<1.2 \times 10^9$~K are not drawn in the
    top and bottom panels, respectively. Two curves in top and bottom
    panels indicate contours of $\tccr/\tdy$ and $\tccs/\tdy$,
    respectively.}
  \label{fig:hotspot_ccdy}
\end{figure*}

We search for hotspots in our simulation at the time when any particle
has the density of $> 2 \times 10^6$~$\gcm$ and the raw temperature of
$>2.5 \times 10^9$~K, and at the time of the first peak (see
section~\ref{sec:overview}). The former time is the same as the time
of the creation of the hotspots defined by \cite{Pakmor12a}, which is
based on results of \cite{Seitenzahl09}. We call this time ``Pakmor's
time''. In Figure~\ref{fig:hotspot_ccdy}, we show the densities and
temperatures of particles at Pakmor's time (left) and the time of the
first peak (right) in model \mdlx{11}. Curves in the top and bottom
panels indicate contours of $\tcc/\tdy$. Note that the criterion of
\cite{Pakmor12a} are stronger than even $\tccr < 0.1\tdy$, since it
considers the decay of the carbon detonation. Although we use $\tccr <
\tdy$ for the criterion of the hotspots below, we also consider
whether the hotspots appear if we choose the constraints of
\cite{Seitenzahl09}.

At Pakmor's time, there are particles satisfying $\tccr < \tdy$ (the
top left panel), but none with $\tccs > \tdy$ (the bottom left panel),
where $\tccr$ and $\tccs$ are evaluated using the raw and smoothed
temperatures, respectively. On the other hand, at the time of the
first peak, $\tccr < \tdy$ (the top right panel) and $\tccs < \tdy$
(the bottom right panel). Therefore, the creation of the hotspots at
Pakmor's time depends on the numerical treatment of the temperature,
while at the time of the first peak the hotspots are created robustly
irrespective of the treatment of temperature in an SPH simulation. We
note that, if we use the criterion of \cite{Pakmor12a} for the
creation of the hotspots, the hotspots are not created even at the
first peak from the point of view of the smoothed temperature.

In our simulation, the merger time is relatively shorter than previous
studies. However, the time to the merger does not affect the peak
temperature throughout the merging process
\citep{Pakmor12b}. Therefore, the above discussion whether the
hotspots are created is robust.

\begin{figure*}
  \begin{center}
    \includegraphics[width=16.0cm]{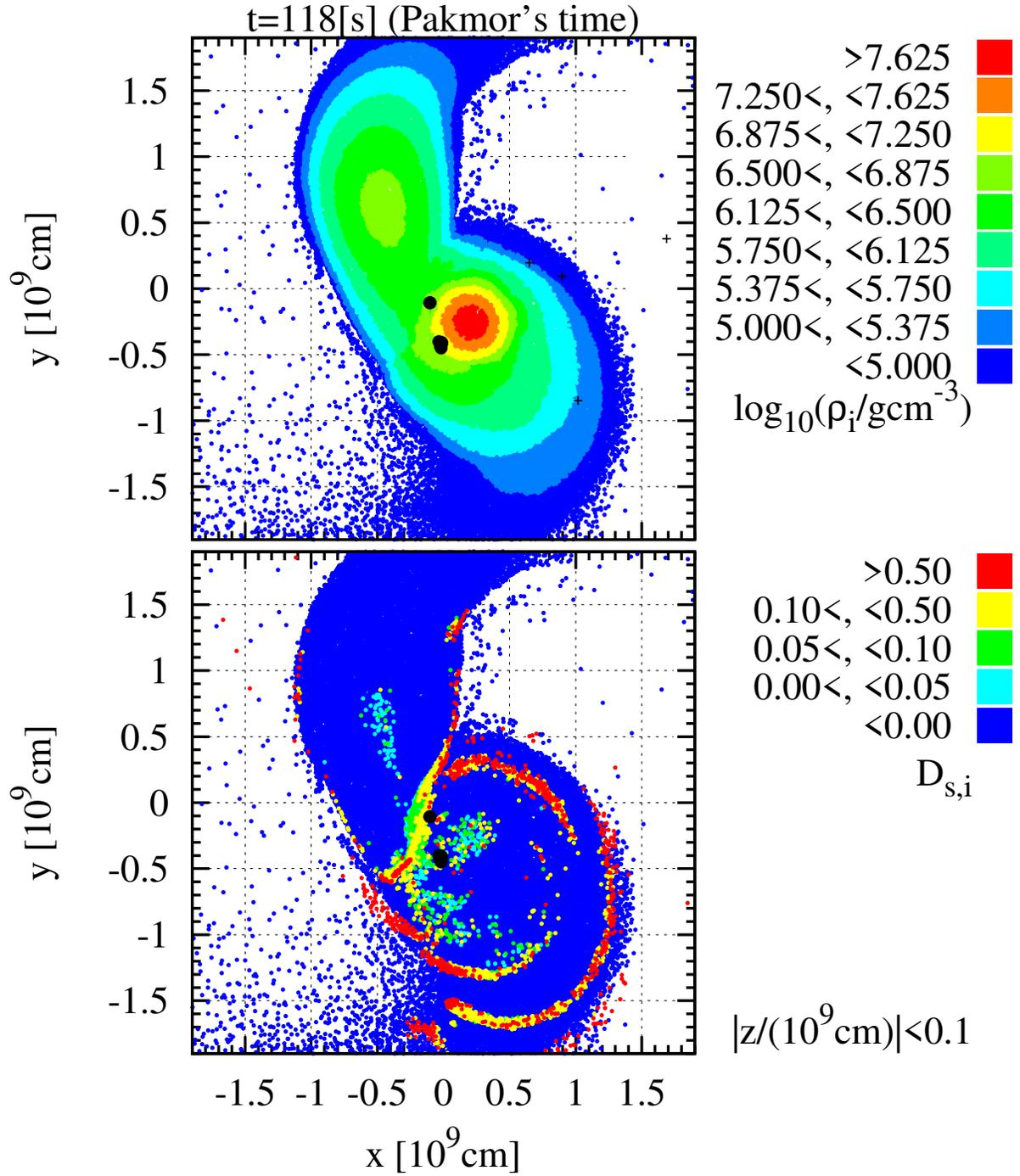}
  \end{center}
  \caption{Mass density (top) and shock detector (bottom) of particles
    at $118$~s in model \mdlx{5.5}. The coordinate origin is at the
    center of mass of the binary system. The particles are separated
    from the orbital plane ($x$--$y$ plane) by $< 0.1 \times
    10^9$~cm. Black dots indicate particles with $\rho_i > 2 \times
    10^6~\gcm$ and $T_{{\rm r},i}>2.5 \times 10^9$~K. In the top
    panel, black crosses indicate the helium particles when $f_{\rm
      He} = 4 \times 10^{-5}$.}
    \label{fig:hotspot_hstr}
\end{figure*}

Hereafter we adopt model \mdlx{5.5} to investigate the formation
mechanism of these hotspots. In model \mdlx{5.5}, Pakmor's time is
$t=118$~s as shown in Figure~\ref{fig:hotspot_hstr}.  Black dots
indicate the hotspots. The hotspots are formed as follows. Just before
$t=118$~s, the companion is tidally disrupted. Subsequently, a large
amount of the disrupted debris is rapidly accreted onto the
primary. This can be seen in the top panel. The center of the primary
is located on the red region, and the debris extends in the direction
of the top left from the primary in this figure. The accretion of the
debris forms a shocked region at the surface of the primary. The
shocked region can be seen from
$(x/10^9\mbox{cm},y/10^9\mbox{cm})=(-0.5, -0.5)$ to $(0, 0.5)$ in the
bottom panel. In this region, the materials are compressed, and the
resulting high raw temperatures lead to the creation of the
hotspots.

This behavior is qualitatively in agreement with the results by
\cite{Pakmor12a} where a system of binary CO~WDs with masses of $1.1$
and $0.9M_{\odot}$, similar to our case, is considered. As seen in
their figure~1, their hotspots are generated at the primary's surface at
the time just after the companion is tidally disrupted.

In order to investigate the effects of the alpha process, we draw the
positions of the helium particles when $f_{\rm He} = 4 \times
10^{-5}$. The reason why we choose $f_{\rm He} = 4 \times 10^{-5}$ is
that the binary explodes in the case of $f_{\rm He} \gtrsim 3 \times
10^{-4}$ before this time through the helium-ignited violent merger
mode (see section~\ref{sec:double}).  As seen in the top panel of
Figure~\ref{fig:hotspot_hstr}, the helium particles are far away from
the hot particles indicated by the black dots. The alpha process does
not affect the carbon-ignited violent merger mode at this time.

Next, we investigate the creation of the hotspots at the time of the
first peak.  Actually, at the time of the first peak in model
\mdlx{5.5}, even the particles with the highest temperature does not
satisfy the necessary condition to form the hotspots if we adopt the
smoothed temperature. Despite of the absence of the hotspots in model
\mdlx{5.5} (for the smoothed temperature), we adopt this model for
further investigation for the following reasons: We unfortunately do
not follow the evolution of model \mdlx{11} until its merger remnant
reaches a dynamically steady state, since the simulation of model
\mdlx{11} is highly time-consuming. As a result, we can not assess the
Chandrasekhar model and investigate merger ejecta, using model
\mdlx{11}. In order to assess all the explosion models with the same
simulation model, we assess the carbon-ignited violent merger model,
using model \mdlx{5.5}. Indeed, the particles in model \mdlx{5.5}
obtain high smoothed temperatures through the same mechanism as the
hotspots found in model \mdlx{11}. Regarding these particles with the
high smoothed temperature as the hotspots (which should satisfy the
condition for the hotspots in the corresponding higher-resolution
simulation), we investigate the formation mechanism of these
particles.

\begin{figure*}
  \begin{center}
    \includegraphics[width=16.0cm]{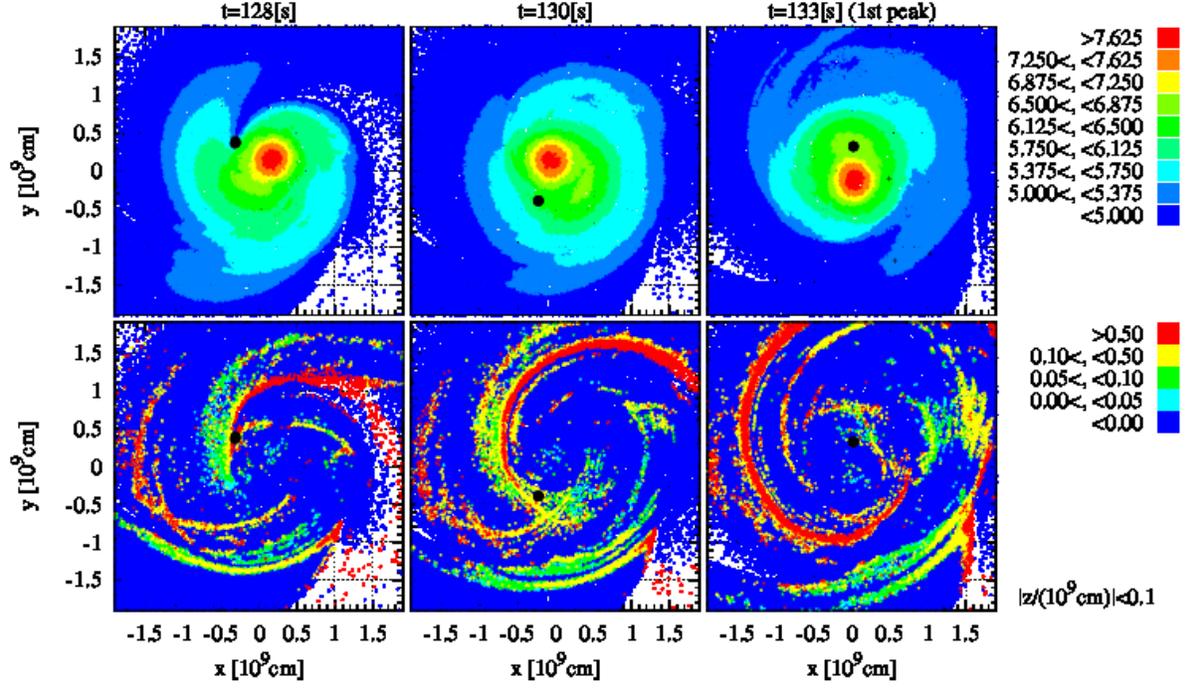}
  \end{center}
  \caption{Mass densities (top) and shock detectors (bottom) of
    particles at $t=128$~s, $130$~s, and $133$~s from left to right in
    model \mdlx{5.5}. The particles are separated from the orbital
    plane by $< 0.1 \times 10^9$~cm. They are colored in the same way
    as Figure~\ref{fig:hotspot_hstr}. Black dots indicate particles
    with smoothed temperatures more than $1.5 \times 10^9$~K at
    $t=133$~s. The numbers of these particles are $4$. In each panel,
    black dots look like only one dot. This is because these particles
    are located on almost the same positions. In the top right panel,
    black crosses indicate the helium particles when $f_{\rm He} = 4
    \times 10^{-5}$.}
  \label{fig:hotspot_hsts1_view}
\end{figure*}

Figure~\ref{fig:hotspot_hsts1_view} shows the states of particles
around at the time of the first peak. The black dots indicate
particles with the smoothed temperature exceeding $1.5 \times 10^9$~K
at $t=133$~s. The number of these particles is $4$. We regard these
particles at $t=133$~s as the hotspots. Note that these particles are
different from those regarded as the hotspots at Pakmor's time.

Similarly to Pakmor's time (see Figure~\ref{fig:hotspot_hstr}), we
draw the helium particles ($f_{\rm He}=4 \times 10^{-5}$) on the top
right panel of Figure~\ref{fig:hotspot_hsts1_view}. The reason why we
choose $f_{\rm He}=4 \times 10^{-5}$ is the same as the case of
Pakmor's time. Similarly to Pakmor's time, the helium particles are
far away from the hot particles indicated by the black
dots. Therefore, the alpha process does not affect the carbon-ignited
violent merger mode at this time.

We follow trajectories of these particles. At $t=128$~s, they are
caught between the primary and a tidal tail. A collision between the
primary and the tidal tail forms a shocked region. When the particles
pass across the shocked region, their kinetic energies are converted
to the internal energies. After $t=128$~s, these particles orbit
around the primary. At $t=133$~s, these particles irrupt into a clump
created by a debris of the tidally disrupted companion. The clump has
a high density, $10^{6.5}<\rho_i/(\gcm)<10^{6.875}$, despite that it
is separated from the center of the primary by $\sim 0.5 \times
10^9$~cm. Its surroundings have a lower density,
$10^{6}<\rho_i/(\gcm)<10^{6.5}$.  When the particles irrupt into the
clump, they are compressed (nearly adiabatically), and achieve the
highest smoothed temperatures.

\begin{figure*}
  \begin{center}
    \includegraphics[width=12.0cm]{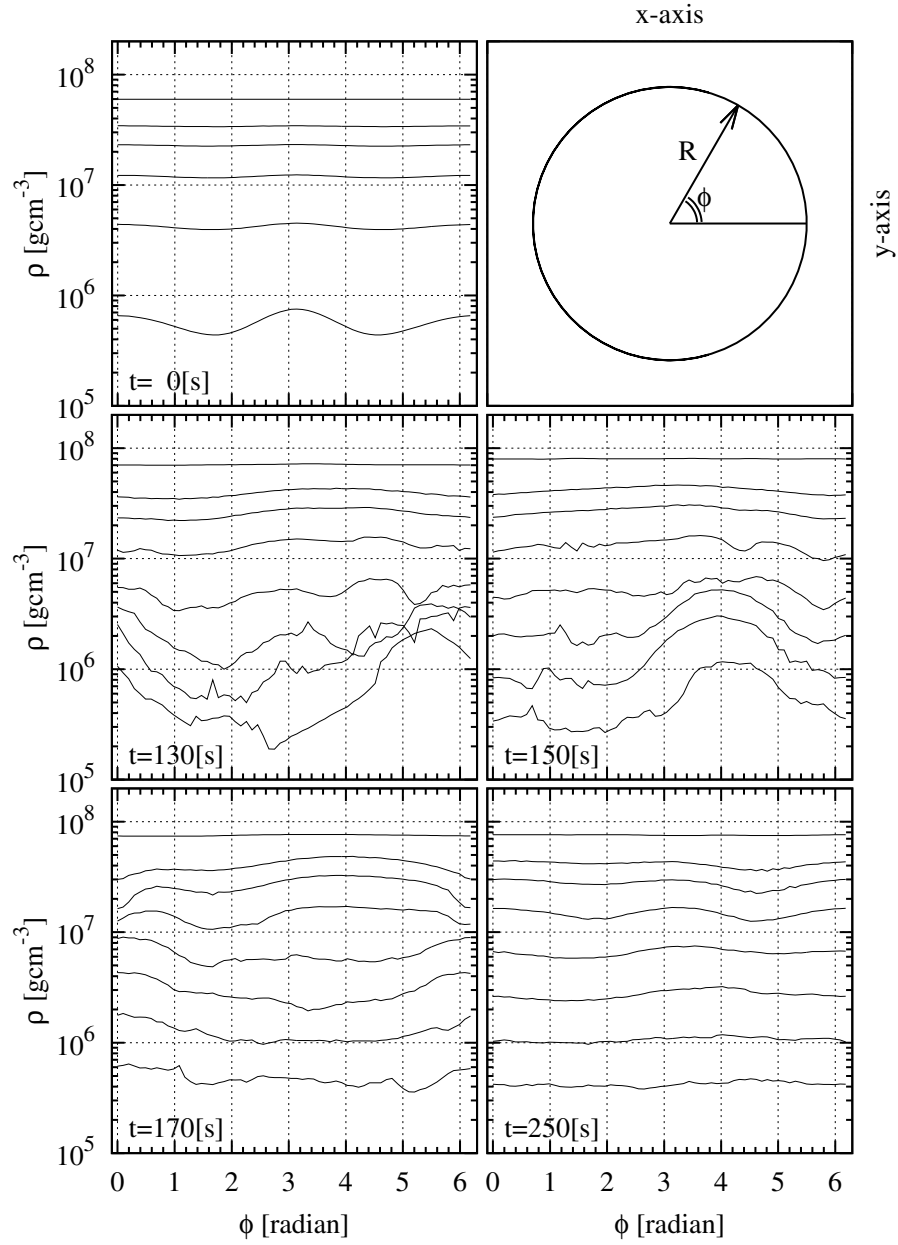}
  \end{center}
  \caption{Mass density on a circle on the orbital plane. The circle
    is centered on the center of the primary. As seen in the top right
    panel, a position on the circle is depicted as a radius $R$ and
    angle $\phi$, where the origin is the center of the primary. In
    the other panels, mass densities are shown on circles with
    $R=0.010$, $0.018$, $0.032$, $0.056$, $0.10$, $0.18$, $0.32$,
    $0.56$ in the units of $10^9$~cm from top to bottom at $t=0$~s,
    $130$~s, $150$~s, $170$~s, and $250$~s.}
  \label{fig:hotspot_asym}
\end{figure*}

Figure~\ref{fig:hotspot_asym} shows the creation and evolution of the
clump quantitatively. The horizontal axes in all but the top right
panel indicate $\phi$, which is an angle between a line segment
connecting the coordinate origin and a given point, and one connecting
the coordinate origin and the initial position of the center of mass
of the primary. At $t=0$~s, the densities are almost independent of
$\phi$ at density exceeding $10^6~\gcm$. However, it is not the case
at $t=130$~s, $150$~s, and $170$~s, during several tens of seconds
after the merger time. The clump is present at $\phi=5$ ($t=130$~s),
$4$ ($t=150$~s), and $0$ radian ($t=170$~s). At $t=250$~s, the density
becomes independent of $\phi$ again, announcing that the clump has
disappeared.

\cite{Kashyap15} have found a hotspot formed through a spiral mode
instability in the accretion disk consisting of the debris of the
companion.  Such a hotspot and spiral possibly appear in our
simulation. We can see a spiral in the bottom panels of
Figure~\ref{fig:hotspot_hsts1_view}. Also, a particle with the highest
temperature at the second peak in our simulation are similar to the
hotspot in \cite{Kashyap15}; our particle has density of $2 \times
10^7$~$\gcm$ and temperature of $2.1 \times 10^9$~K in \mdlx{5.5} (see
Table~\ref{tab:maximum_temperature}), while the hotspot has density of
$10^7$~$\gcm$ and temperature of $3 \times 10^9$~K. Our particle has
slightly smaller temperature than the hotspot, since nuclear reactions
are not solved in our simulation. We do not discuss the hotspot (or
the particle with the highest temperature at the second peak) anymore.

Since we consider only the necessary condition to lead to the
initiation of the carbon detonation and do not deal with subsequent
evolution following the detonation, it is not clear whether this
binary system explodes in the end or not. Also, it is uncertain
whether the explosion is initiated at Pakmor's time or at the time of
the first peak -- both are possible but our understanding is currently
limited by numerical difficulties.  Nevertheless, we investigate what
the explosion should look like as we did for the helium-ignited
violent merger model, especially considering two situation where the
explosion occurs either at Pakmor's time or at the time of the first
peak. We again focus on the expected early brightness and \fsni
distribution within the hypothesized SN ejecta.

\begin{figure*}
  \begin{center}
    \includegraphics[width=16.0cm]{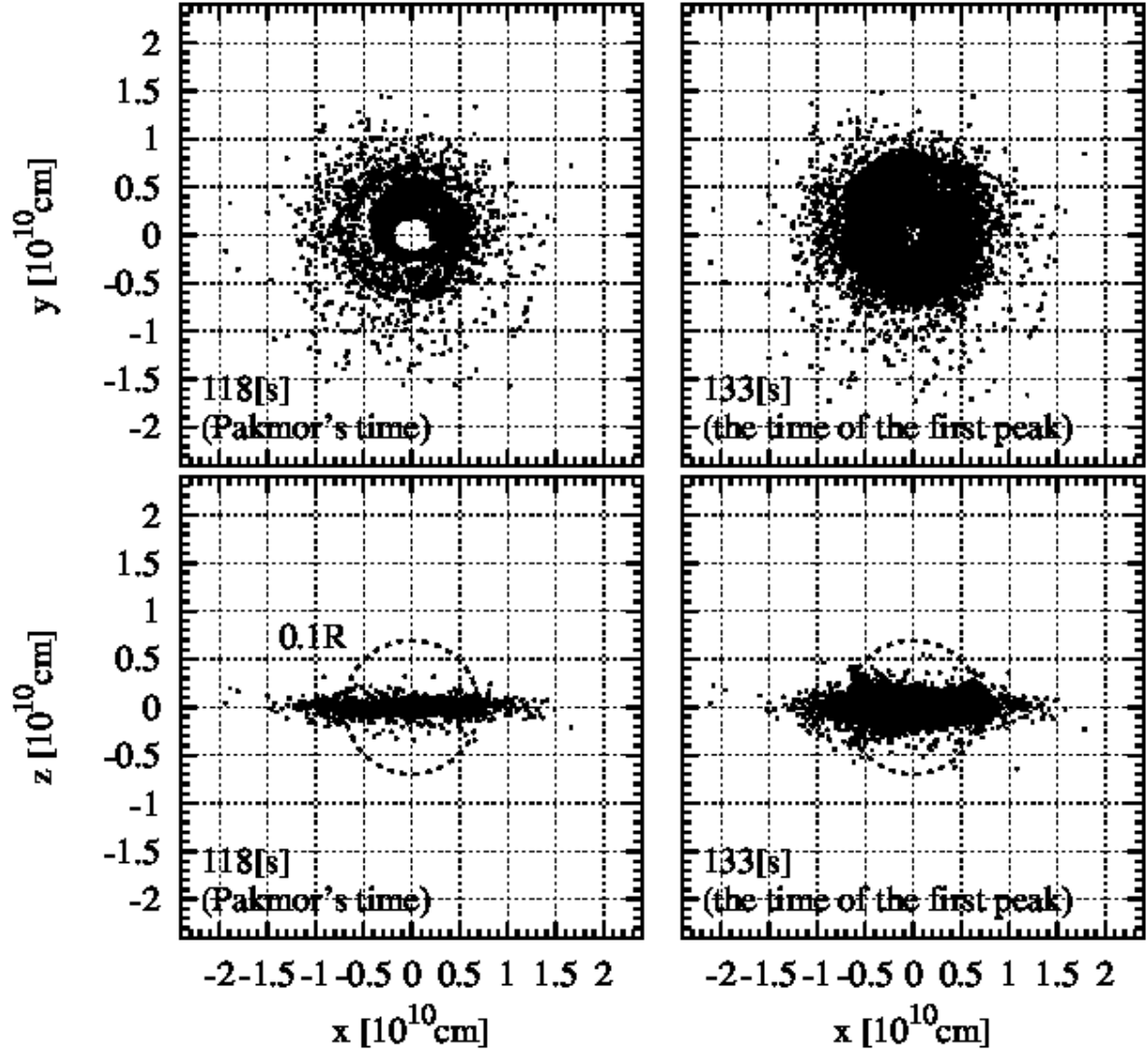}
  \end{center}
  \caption{Material distribution of a binary with $1.1M_\odot$ and
    $1.0M_\odot$ in model \mdlx{5.5}. We draw only particles separated
    from the center of the primary by $2 \times 10^9$~cm. The time is
    indicated at the bottom left in each panel. The time is the
    initiation time of the carbon detonation in the cases of the raw
    temperature (two left panels) and the smoothed temperature (two
    right panels). Dashed curves indicate $0.1R_\odot$ from the center
    of the primary.}
  \label{fig:debris_sdis_carbon}
\end{figure*}

Figure~\ref{fig:debris_sdis_carbon} shows the material distribution of
our binary at Pakmor's time and at the time of the first peak. The
materials spread out beyond $0.1R_\odot$, and reach up to $\sim
0.3R_\odot$. The system has a disk structure, similar to the material
just before an explosion in the helium-ignited violent merger model
(see section~\ref{sec:double}). However, the disk in this case is more
massive and thicker than in the case of the helium-ignited violent
merger model.

Following the same argument as presented in section~\ref{sec:double}
but applied to the expected pre-SN structure for the carbon-ignited
violent merger model, we expect that this mode results in the early
brightness much brighter than that of SN~2011fe, and much fainter than
that of SN~2014J.  This is because the envelope has a radius of $\sim
0.3R_\odot$, which is much larger than that inferred for SN~2011fe ($<
0.1R_\odot$), but much smaller than that for SN~2014J ($>
1R_\odot$). The situation is similar to that for the helium-ignited
violent merger model, and again our comparison suffers from the
limitation of the spherically symmetric structure assumed in the
estimates of the progenitor radii for these SNe.  The envelope has
disk-like structure (see Figure~\ref{fig:debris_sdis_carbon}), and
this should be taken into account for detailed
comparison. Qualitatively, the small opening angle of the envelope
from the primary should make the early brightness fainter than we
expected above, and the light curve may be consistent with that of
SN~2011fe. This should be quantitatively investigated by means of
numerical simulations, similarly to the early brightness resulting
from the helium-ignited violent merger model (see
section~\ref{sec:double}).

Next we discuss \fsni distribution synthesized at the hypothesized
explosion.  If the explosion is initiated at Pakmor's time, its
distribution is similar to that of \cite{Pakmor12a}. Briefly speaking,
\fsni is expected to be absent in the central region of the SN ejecta
in the following reason: Since the companion is burned later than the
primary, the ashes of the companion are expected to be located at the
central region in the SN ejecta. At the same time, the low density
there results in little amount of $^{56}$Ni. On the other hand, if the
explosion is initiated at the time of the first peak, \fsni is present
at the central region of the explosion in the following reason.  Since
the companion has been largely disrupted already, the system is more
spherically symmetric in this case than at Pakmor's time. The hotspots
are created at the surface of the primary, and the explosion is
initiated at the off-center region. This configuration is similar to
that in gravitationally confined detonation model
\citep{Jordan08,Meakin09} or the off-center delayed-detonation model
\citep{Kasen09,Maeda10,Seitenzahl13}, in which a large amount of
$^{56}$Ni is synthesized near the center of the SN ejecta.  The
difference of \fsni distribution comes from whether the companion is
disrupted or not at the time of the explosion.

From an observational view point, a model where the explosion is
initiated at the time of the first peak is more favorable, since SN~Ia
contains \fsni (or other Fe-peak elements) near the center of the
explosion.  However, even in this case, after the explosion the \fsni
distribution is expected to evolve to an hourglass-like shape, since
\fsni is prevented from moving toward the direction of the orbital
plane as blocked by a debris of the companion \citep{Raskin14}. This
is not consistent with the \fsni distribution generally inferred for
SN~Ia which is a spherically symmetric shape \citep{Maund13,Soker14}.

This section is summarized as follows: In the carbon-ignited violent
merger model, we confirmed that the hotspots appear. Therefore, an
explosion can potentially occur through this mode.  However, it is
expected that the resulting early brightness is much brighter than
SN~2011fe and much fainter than SN~2014J. Note that the small opening
angle of the envelope from the primary should make the light curve
fainter than we expect, and that the light curve may be consistent
with that of SN~2011fe.  Moreover, it is expected that the \fsni
distribution does not have a spherically symmetric shape, also being
inconsistent with \fsni distribution generally inferred for SN~Ia. The
expected observational outcome will apply to any explosions through
this mode, since the explosions can potentially occur only from the WD
mass combinations similar to one studied in this paper \citep{Sato15}.

\subsection{Other models}
\label{sec:other}

An explosion in the carbon-ignited violent merger model happens around
the merger time, while an explosion in the Chandrasekhar mass model
happens $\sim 10^4$ years after the merger time
\citep[e.g.][]{Yoon07}. Between these two epochs, an explosion through
other models possibly happens, given that the system does not
experience the explosion through the helium- and carbon-ignited
violent merger models.  For example, \cite{Schwab12} and \cite{Ji13}
have suggested an explosion triggered by magnetohydrodynamical effects
in this phase.  In this section, we do not assess whether the
explosion happens in such a model as it is beyond what we can discuss
based on our pure hydrodynamic simulation, but discuss what the
explosion should look like, assuming that the explosion does happen.

\begin{figure*}
  \begin{center}
    \includegraphics[width=12.0cm]{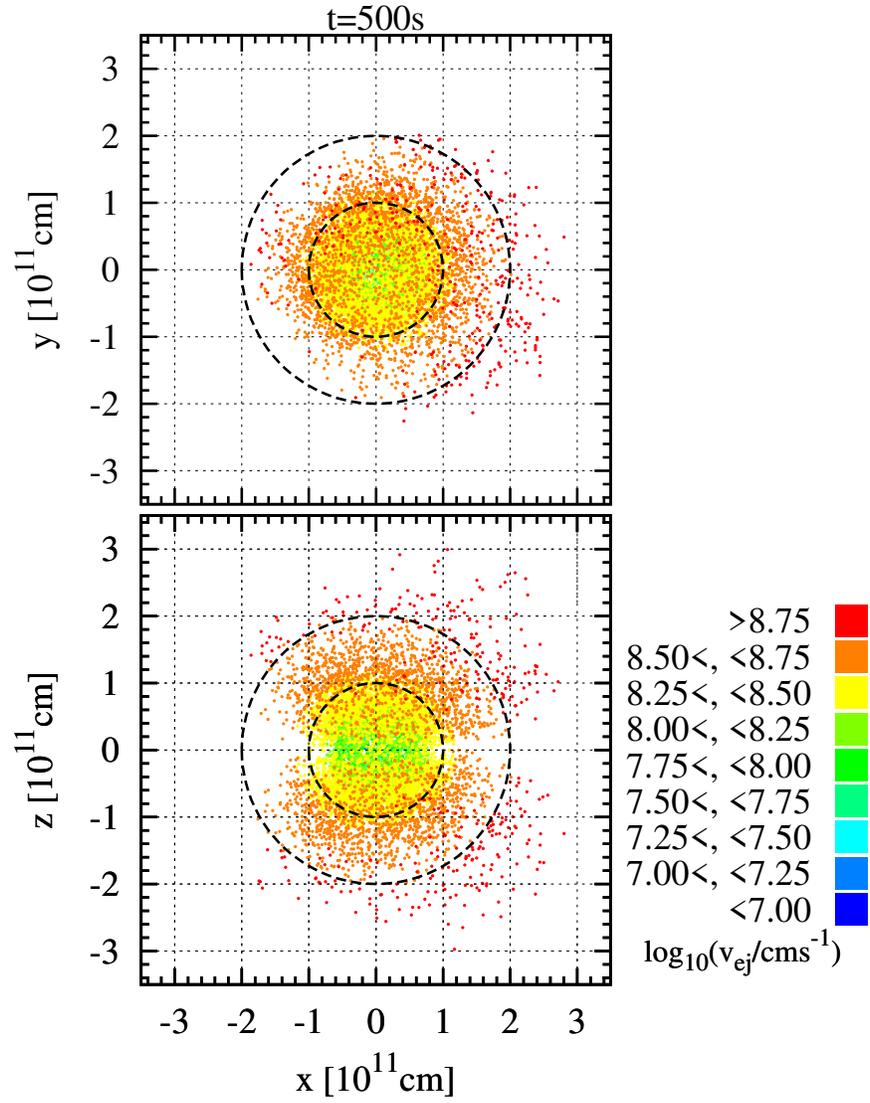}
  \end{center}
  \caption{Distribution of merger ejecta at $t=500$~s in model
    \mdlx{5.5}. The ejecta are colored according to their $v_{\rm
      ej}$, defined in equation~(\ref{eq:vej}). All ejecta are
    projected on the orbital and $x$--$z$ planes in the top and bottom
    panels, respectively.}
  \label{fig:ejecta_sdis}
\end{figure*}

The appearance of the explosion will be affected by the nature of
merger ejecta. Figure~\ref{fig:ejecta_sdis} shows the spatial and
velocity distributions of the merger ejecta at $500$~s. The merger
ejecta spread almost isotropically, except that they are relatively
deficient on the orbital plane (see also
Figure~\ref{fig:ejecta_shck}).  Nevertheless, the covering factor of
the merger ejecta around the merger remnant is almost unity.

\cite{Raskin13} have also studied effects of merger ejecta on
observations of a putative SN taking place in this phase/mode.
However, they have focused only on the tidal ejecta for the merger of
binary CO~WDs with masses of $0.96M_{\odot}$ and $0.64M_{\odot}$. On
the other hand, our binary model consists of $1.1M_\odot$ and
$1.0M_\odot$ CO~WDs. Since our binary has more massive components and
a mass ratio closer to unity, our binary merges more violently than
theirs \citep{Marsh04}. Consequently, our merger ejecta are dominated
by the shocked ejecta (see section~\ref{sec:overview}). Therefore, the
expected effects of the merger ejecta on the observational features
are partly different from those described by \cite{Raskin13}.

\cite{Raskin13} have shown that NaID absorption features are
potentially observed with a probability of $10$ -- $50$ percent if the
explosion occurs $10^8$~s -- $10^2$~yr or $10^3$~yr -- $10^5$~yr after
the merger time. The probability corresponds to the covering factor of
their merger ejecta. However, in the case of our binary, the NaID
absorption features are potentially observed in almost all the
cases.\footnote{Adding to this, one has to consider the thermal
  condition of the ejecta, which is beyond the scope of this paper.}
This result is complementary with the discussion in section~3.4 and
3.6 of \cite{Raskin13}.

\begin{figure*}
  \begin{center}
    \includegraphics[width=12.0cm]{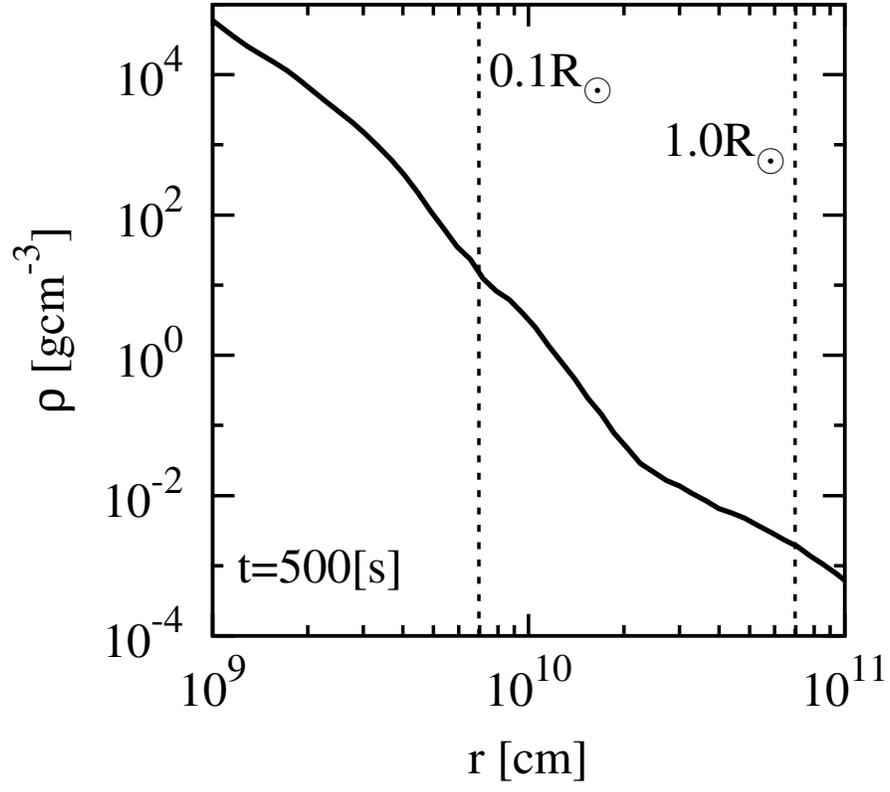}
  \end{center}
  \caption{Distribution of the mass density of particles in model
    \mdlx{5.5} at $t=500$~s. They are mapped into a one-dimensional
    profile; the horizontal axis indicates the spherical radius. The
    vertical dotted lines indicate a distance of $0.1R_{\odot}$ and
    $1.0R_{\odot}$ from the center of the merger remnant.}
  \label{fig:debris_rho}
\end{figure*}

Finally, we point out that, if an explosion does happen at $t=500$~s,
the explosion might look like SN~2014J in the early brightness.  At
$t=500$~s, the envelope of our binary model spreads out beyond
$1.0R_{\odot}$, as seen in Figure~\ref{fig:debris_rho}. The horizontal
axis indicates the spherical radius.

\subsection{Chandrasekhar mass model}
\label{sec:chandrasekhar}

In this section, we explore a possibility of an explosion in the
Chandrasekhar mass model, which might take place if the system does
not undergo an explosion through the mechanisms investigated by the
previous sections.  After the remnant reaches a dynamically steady
state, it gradually loses the thermal energy by neutrino cooling, and
increases its central density and temperature.  If its central density
and temperature exceed critical values as set by the balance between
the nuclear reaction timescale and dynamical or convection time scale,
the remnant is likely to explode as an SN Ia (`the Chandrasekhar mass
model') assuming the central region consists of carbon.  On the other
hand, it likely collapses to a neutron star as triggered by the
electron capture, if the remnant has become an oxygen-neon-magnesium
WD before reaching to this phase.

The presence of carbon in the remnant depends on whether carbon is
quiescently burned. All of carbon in the remnant are burned, if the
$^{12}$C + $^{12}$C reactions proceed faster than the neutrino cooling
at the time when the remnant reaches a dynamically steady state
\citep{Saio85,Saio98,Saio04}.

\begin{figure*}
  \begin{center}
    \includegraphics[width=16.0cm]{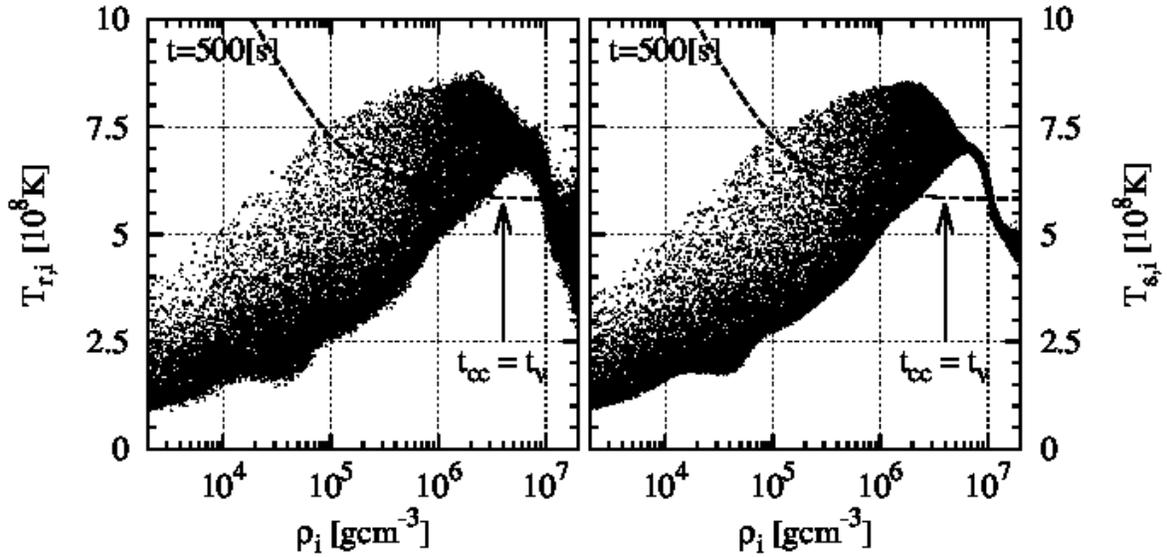}
  \end{center}
  \caption{Mass density and temperature of particles at $t=500$~s in
    model \mdlx{5.5}. The vertical axes indicate raw and smoothed
    temperatures in the left and right panels, respectively. The
    dashed curves indicate contours with $\tcc = t_{\rm \nu}$. Above
    the curves, $\tcc < t_{\rm \nu}$.}
  \label{fig:chandra_rhot}
\end{figure*}

Figure~\ref{fig:chandra_rhot} shows mass density and temperature of
particles at $t=500$~s in model \mdlx{5.5}. At that time, the merger
remnant reaches a dynamically steady state. In many particles, the
timescale of the $^{12}$C + $^{12}$C reaction is shorter than the
timescale of the neutrino cooling, where we calculate the neutrino
cooling rate using a public code available at F.~X.~Timmes
website\footnote{http://cococubed.asu.edu/code\_pages/nuloss.shtml}
which is based on \cite{Itoh96}. However, the timescale of $^{12}$C +
$^{12}$C reaction is longer than the dynamical timescale.  Therefore,
the $^{12}$C + $^{12}$C reaction does not trigger the carbon
detonation, but will convert the merger remnant to an
oxygen-neon-magnesium WD on thermal timescale.  We conclude that,
after the remnant evolves on thermal timescale, the remnant would not
explode in the Chandrasekhar mass model, rather collapses to a neutron
star.

We should keep in mind that this result may be affected by the short
merger time in our simulation. According to \cite{Dan11}, the maximum
temperature in the merger remnant becomes high when the binary
suddenly merges. Therefore, if the time to the merger is longer, the
maximum temperature in the merger remnant may be lower, and the
$^{12}$C + $^{12}$C reaction rate is lower than the neutrino cooling
rate. Then, the remnant could explode in the Chandrasekhar mass model.

\section{Detectability of merger remnant and merger shell}
\label{sec:detectability}

If binary CO~WDs (or its merger remnant) fail to explode in all but
the Chandrasekhar mass model, the merger ejecta have at least
$10^4$~yr of time during which the ejecta expand into the ISM.  An
expanding shell is formed as the merger ejecta sweep up its
surrounding ISM, analogous to an SN remnant.  Also, the merger remnant
still exists during this phase, analogous to a neutron star in a
core-collapse SN remnant.  For this situation, we estimate a
detectability of the merger remnant and merger shell. In particular,
we focus on the merger shell, estimating its luminosity.

The evolution of a merger shell is divided into three phases.  The
first phase is a `free expansion phase', where the amount of ISM swept
up by the merger ejecta is negligible. Once the shell sweeps up ISM
mass comparable to the mass of the merger ejecta, the shell is
substantially decelerated entering into the `Sedov phase'. Materials
are thermalized behind the shock wave. Because of the high
temperature, the shell loses only a negligible fraction of energy
through radiation approximately conserving the total energy content.
Once the shell is cooled down, a significant amount of energy is lost
by radiation, where the momentum is approximately conserved.  This
phase is called a `snowplow phase'.

A luminosity of the shell is notated by $L_{\rm shell}$, and given by
\begin{align}
  L_{\rm shell} = \Lambda V_{\rm shell}, \label{eq:l1}
\end{align}
where $\Lambda$ is a cooling function, and $V_{\rm shell}$ is the
volume of the shell. The volume $V_{\rm shell}$ can be expressed as
$V_{\rm shell}=4\pi R_{\rm s}^2\Delta_{\rm shell}$, where $R_{\rm s}$
is a radius of the shock wave in front of the shell, and $\Delta_{\rm
  shell}$ is the thickness of the shell. We give the cooling function
$\Lambda$ as follow:
\begin{align}
  \Lambda = \beta T_{\rm shell}^{-0.7} n_{\rm e, shell} n_{\rm H,
    shell},
\end{align}
where $n_{\rm e,shell}$ and $n_{\rm H,shell}$ are the number densities
of electrons and hydrogen atoms in the shell, $T_{\rm shell}$ is the
temperature in the shell, and $\beta=1.7 \times
10^{-18}$~ergcm$^{-3}$s$^{-1}$K$^{0.7}$ \citep{Draine11}. Note that
this cooling function takes into account metal emission lines, and is
applicable in the range of $10^{5} < T_{\rm shell}/\mbox{K} <
10^{7.3}$. Since hydrogen atoms are perfectly ionized after they pass
through the shock wave, the number densities $n_{\rm e,shell}$ and
$n_{\rm H,shell}$ can be expressed as $n_{\rm H,ism}
(\gamma+1)/(\gamma-1)$.

We estimate the shell luminosity at the free expansion. At the free
expansion phase, the radius of the shock wave $R_{\rm s} \propto t$,
and the thickness $\dot{R}_{\rm s}$ is constant. Then, $V_{\rm shell}
\propto t^3$, and the temperature in the shell, $T_{\rm shell}$, is
constant. Eventually, $L_{\rm shell} \propto t^3$.

We focus on the Sedov phase, since the luminosity $L_{\rm shell}$
reaches to the peak around the end of the Sedov phase (see
below). Although, at the free expansion phase, we show only the
proportional relation between $L_{\rm shell}$ and $t$, we give the
equation expressing the relationship between $L_{\rm shell}$ and $t$
at the Sedov phase. Assuming a strong shock wave, the radius ($R_{\rm
  s}$) and the thickness ($\Delta_{\rm shell}$) can be written as
\begin{align}
  R_{\rm s} &= \left[ \frac{75}{16\pi}
    \frac{(\gamma-1)(\gamma+1)^2}{(3\gamma - 1)} \frac{E_{\rm
        k,ej}}{m_{\rm H} n_{\rm ism}} \right]^{0.2}
  t^{0.4} \label{eq:rs} \\
  \Delta_{\rm shell} &= \frac{(\gamma - 1)}{3(\gamma + 1)} R_{\rm s},
\end{align}
where $E_{\rm k,ej}$ is the total kinetic energy of the merger ejecta,
$n_{\rm ism}$ is the number density of the ISM in front of the shock
wave, $m_{\rm H}$ is the mass of a hydrogen atom, and $\gamma$ is the
adiabatic index \citep{Cavaliere76}. Note that we assume that the ISM
consists only of hydrogen atom here to provide a first order estimate.

The shell temperature $T_{\rm shell}$ can be
obtained as follows. From Rankine-Hugoniot conditions, the shell
pressure $p_{\rm shell}$ is given by
\begin{align}
  p_{\rm shell} = m_{\rm H}n_{\rm ism}\dot{R}_{\rm
    s}^2/(\gamma+1), \label{eq:p1}
\end{align}
where $\dot{R}_{\rm s}$ is the speed of the shock wave. On the other
hand, the pressure $p_{\rm shell}$ can be written from an equation of
state of an ideal gas as
\begin{align}
  p_{\rm shell}=n_{\rm H,shell}k_{\rm B}T_{\rm shell}, \label{eq:p2}
\end{align}
where $k_{\rm B}$ is the Boltzmann constant. Using
equation~(\ref{eq:p1}) and (\ref{eq:p2}), we obtain the shell
temperature:
\begin{align}
  T_{\rm shell} = \frac{2(\gamma-1)}{(\gamma+1)^2} m_{\rm H} k_{\rm
    B}^{-1} \dot{R}_{\rm s}^2. \label{eq:tsh}
\end{align}
At the Sedov phase, the speed of the shock wave $\dot{R}_{\rm s}$ can
be written as
\begin{align}
  \dot{R}_{\rm s} = \left[ \frac{3}{4\pi}
  \frac{(\gamma-1)(\gamma+1)^2}{(3\gamma-1)} \frac{E_{\rm
      k,ej}}{m_{\rm H}n_{\rm ism}} \right]^{0.5} R_{\rm s}^{-1.5}
\end{align}
\citep{Cavaliere76}.

Then, we can rewrite equations~(\ref{eq:l1}) and (\ref{eq:rs}) as
\begin{align}
  L_{\rm shell} &= 2.6 \times 10^{36} \left( \frac{n_{\rm
      ism}}{1\mbox{cm$^{-3}$}} \right)^{1.68} \nonumber \\ 
  &\times \left( \frac{E_{\rm k,ej}}{3.2 \times 10^{47}\mbox{erg}}
  \right)^{0.32} \left( \frac{t}{10^4 \mbox{yr}} \right)^{2.04}
        [\ergs], \label{eq:l2} \\
  R_{\rm s} &= 2.6 \left( \frac{n_{\rm ism}}{1\mbox{cm}^{-3}}
  \right)^{-0.2} \nonumber \\ &\times \left( \frac{E_{\rm k,ej}}{3.2
    \times 10^{47}\mbox{erg}} \right)^{0.2} \left( \frac{t}{10^4
    \mbox{yr}} \right)^{0.4} \mbox{[pc]}. \label{eq:rs2}
\end{align}
where we set $\gamma = 5/3$. As seen in the power of $t$ in
equation~(\ref{eq:l2}), the shell luminosity keeps increasing as time
goes by. We estimate roughly the time when the Sedov phase is
terminated, $t_{\rm cool}$, using the shell luminosity $L_{\rm shell}$
and the kinetic energy of the merger ejecta $E_{\rm k,ej}$ as:
\begin{align}
  E_{\rm k,ej} = \int_{0}^{t_{\rm cool}} L_{\rm shell}
  dt. \label{eq:tcool1}
\end{align}
Solving equation~(\ref{eq:tcool1}), we obtain $t_{\rm cool}$
as follows:
\begin{align}
  t_{\rm cool} = 1.1 \times 10^4 &\left( \frac{n_{\rm
      ism}}{1\mbox{cm$^{-3}$}} \right)^{0.553} \nonumber \\
  &\times \left( \frac{E_{\rm k,ej}}{3.2 \times 10^{47}\mbox{erg}}
  \right)^{0.224} \mbox{ [yr]}. \label{eq:tcool2}
\end{align}

The cooling function $\Lambda$ is appropriate only when $10^{5} <
T_{\rm shell}/\mbox{K} < 10^{7.3}$. From equation~(\ref{eq:tsh}), we
obtain the following expression:
\begin{align}
  T_{\rm shell} &= 2.3 \times 10^5 \left( \frac{n_{\rm
      ism}}{1\mbox{cm$^{-3}$}} \right)^{-0.4} \nonumber \\
  &\times \left( \frac{E_{\rm k,ej}}{3.2 \times 10^{47}\mbox{erg}}
  \right)^{0.4} \left( \frac{t}{10^4 \mbox{yr}} \right)^{-1.2}
  \mbox{[K]}. \label{eq:tsh2}
\end{align}
From equation~(\ref{eq:tsh2}), the cooling function can be applied
from a few $10^2$ yrs to a few $10^4$ yrs.

At the snowplow phase, ISM which passes through the shock wave emits
energy almost instantly. The shell luminosity is written as $L_{\rm
  shell} \propto \dot{m}_{\rm s} e_{\rm s}$, where $\dot{m}_{\rm s}$
is the rate of the ISM mass passing the shock wave, and $e_{\rm s}$ is
the specific energy which the ISM gains from the shock wave. Since
$R_{\rm s} \propto t^{2/7}$ and $\dot{R}_{\rm s} \propto t^{-5/7}$,
$\dot{m}_{\rm s} \propto t^{-1/7}$. From Rankine-Hugoniot conditions,
$e_{\rm s} \propto \dot{R}_{\rm s}^2 \propto t^{-10/7}$. Then, $L_{\rm
  shell} \propto t^{-11/7}$ at the snowplow phase.

\begin{figure*}
  \begin{center}
    \includegraphics[width=12.0cm]{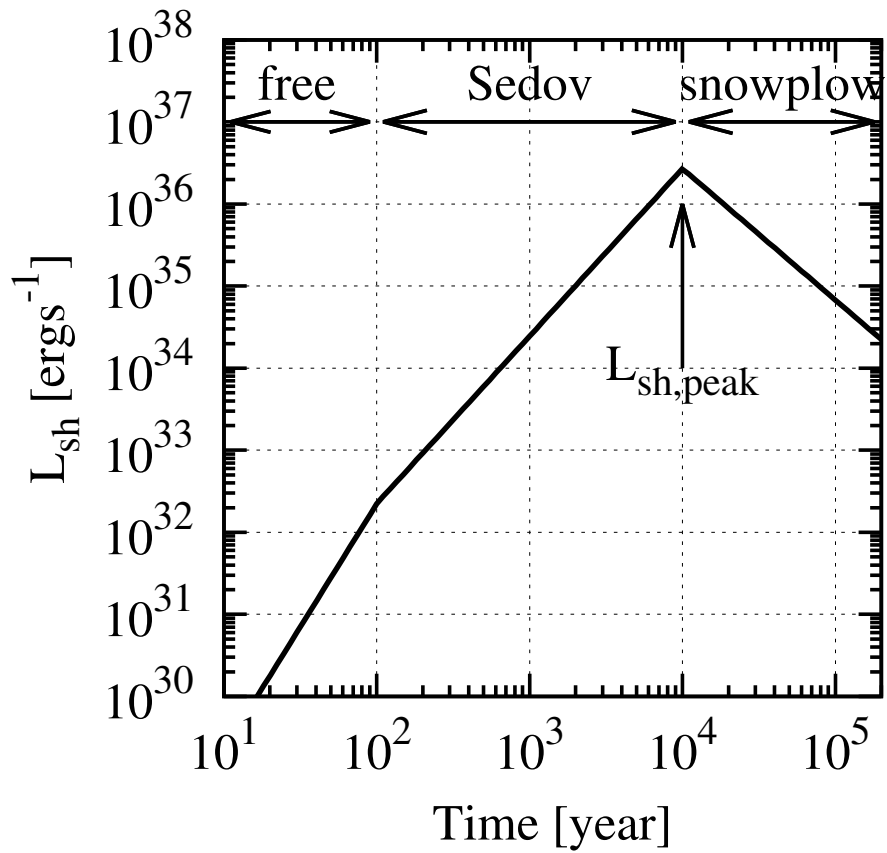}
  \end{center}
  \caption{Time evolution of total luminosity of merger shell at the
    free expansion phase, Sedov phase, and snowplow phase in the case
    of $E_{\rm k,ej}=3.2 \times 10^{47}$~erg.}
  \label{fig:ejecta_shel}
\end{figure*}

In Figure~\ref{fig:ejecta_shel}, we illustrate the time evolution of
the shell luminosity for a set of typical parameters corresponding to
our system; $E_{\rm k,ej}=3.2 \times 10^{47}$~erg, assuming $n_{\rm
  ism}=1$~cm$^{-3}$. We apply equation~(\ref{eq:l2}) for the Sedov
phase. We define the time when the Sedov phase begins as the time when
the shell has swept up the ISM mass comparable to the ejecta mass. The
speed of the shock wave is set to the average velocity of the merger
ejecta, $\sim 3 \times 10^8~\cms$ at the free expansion phase. For the
time when the Sedov phase ends, we adopt $t_{\rm cool}$ in
equation~(\ref{eq:tcool2}).

As seen in Figure~\ref{fig:ejecta_shel}, the shell luminosity reaches
to the peak ($L_{\rm shell,peak}$) at the time when the Sedov phase
ends, $t_{\rm cool}$. Substituting $t_{\rm cool}$ in
equation~(\ref{eq:tcool2}) into $t$ in equation~(\ref{eq:l2}), we
obtain the dependence of the peak luminosity of the shell on the
kinetic energy of the merger ejecta as
\begin{align}
  L_{\rm shell,peak} &= 3.1 \times 10^{36} \left( \frac{n_{\rm
      ism}}{1\mbox{cm$^{-3}$}} \right)^{2.81} \nonumber \\
  &\times \left( \frac{E_{\rm k,ej}}{3.2 \times 10^{47} \mbox{erg}}
  \right)^{0.776} [\ergs].
\end{align}
We define a lifetime of the shell, $T_{\rm life}$, during which the
shell has more than half of the peak luminosity. Then, the lifetime is
given by
\begin{align}
  T_{\rm life} = 4 t_{\rm cool}.
\end{align}

The above estimate on a peak luminosity and lifetime of a merger shell
is based on results of a merger of WDs with masses of $1.1M_\odot$ and
$1.0M_\odot$. However, such a massive binary system is rare.  To
connect the predictions to observations, we therefore scale the above
result to a more common situation of a binary whose total mass just
exceeds the Chandrasekhar mass.  Specifically, we consider a merger of
WDs with masses of $0.9M_{\odot}$ and $0.6M_{\odot}$. This system is
representative of the one in which the merging process proceeds in the
least violent manner amongst the systems potentially leading later to
an explosion through the Chandrasekhar mass model in the ignition
mode. According to our simulation for merging binary CO~WDs with
masses of $0.9$ and $0.6M_\odot$ (see section~\ref{sec:comparison}),
its merger ejecta have the total kinetic energy of $7 \times
10^{45}$~erg.  Consequently, its merger shell will have the peak
luminosity of $L_{\rm shell,peak} = 2 \times 10^{35}$~$\ergs$, and the
lifetime of $T_{\rm life} = 2 \times 10^4$~yr. We regard these values
as the lower limits of the peak luminosity and lifetime of merger
shells, which would be created by any combinations of binary WD masses
if the total mass exceeds the Chandrasekhar mass (hereafter
`super-Chandrasekhar binaries').

The shell could be more luminous than estimated above. This is because
the shell could be illuminated by the merger remnant. Since the merger
remnant has a high temperature for a while after the merger, it would
also have a substantial luminosity. This is analogous to a planetary
nebula. However, we do not consider this effect, but simply note that
our estimate on the merger shell detectability should be regarded as a
lower limit. .

The merger shells will emit ultraviolet and soft X-ray photons.  We
estimate the number of the merger shells in Milky Way, $N_{\rm
  shell}$, assuming that super-Chandrasekhar binaries explode only in
the Chandrasekhar mass model, or fail to explode in all the explosion
models. Then, $N_{\rm shell}$ is given by
\begin{align}
  N_{\rm shell} &\sim 10 \left( \frac{\Gamma_{\rm
      merge}}{10^{-14}\mbox{yr}^{-1}M_{\odot}^{-1}} \right) \nonumber
  \\
  &\times \left( \frac{M_{\rm MW}}{6 \times 10^{10}M_{\odot}} \right)
  \left( \frac{T_{\rm life}}{2 \times 10^4~\mbox{yr}} \right),
\end{align}
where $\Gamma_{\rm merge}$ is a merger rate of super-Chandrasekhar
binaries per unit mass in Milky Way, $M_{\rm MW}$ is the mass of Milky
Way. We adopt $\Gamma_{\rm merge}$ in \cite{Badenes12}, and $M_{\rm
  MW}$ in \cite{McMillan11} and \cite{Licquia14}.

We note difficulties in distinguishing merger shells from SN remnants
and nova shells. Since their explosion energies are different by
several orders of magnitudes, their sizes and luminosities can be
distinguished, if the number density of ISM is known. However, the
number density is usually unknown. One way to overcome this difficulty
to identify the merger shells is to use the information about the
central compact object.

We search for any hint of the merger shell in the literature dealing
with pre-explosion images of SNe~Ia in order to check whether these
SNe~Ia involve merger shells. \cite{Nielsen12} \citep[see
  also][]{Liu12,Nielsen14} have constrained the upper limit of
bolometric luminosities of nearby SNe~Ia with pre-explosion images.
The most stringent limit is on SN~2011fe, $\sim 10^{36}~\ergs$. This
is still lager than the estimated luminosities of merger shells.
Unfortunately, these images are not so deep to constraint the presence
of the merger shells.

As seen in equation~(\ref{eq:rs2}), a merger shell has a parsec-scale
size. When an SN occurs within the merger shell, the merger shell is
not disturbed by the SN ejecta during the first $10^2$~yr after the SN
explosion. Therefore, existence of the merger shell could be also
tested by observations of an SN~Ia after the explosion.  We postpone
such a study to future.

\section{Summary}
\label{sec:summary}

We have performed SPH simulations for merging binary CO~WDs with
masses of $1.1$ and $1.0$ $M_\odot$, until the merger remnant reaches
a dynamically steady state. Using these results, we assess whether the
binary could induce a thermonuclear explosion, and whether the
explosion could be observed as an SN~Ia. We investigate three
explosion mechanisms: a helium-ignition following the dynamical merger
(`helium-ignited violent merger model'), a carbon-ignition
(`carbon-ignited violent merger model'), and an explosion following
the formation of the Chandrasekhar mass WD (`Chandrasekhar mass
model').  In addition to the evaluation if the resulting system
satisfies requirements set in each mode. We have discussed whether the
resulting explosions, through different ignition modes, would look
like SNe~Ia.

Our results are summarized as follows:
\begin{itemize}
\item In the helium-ignited violent merger model, our binary can
  explode, if the mass fraction of helium exceeds a critical value,
  i.e., $\fhe \gtrsim 3 \times 10^{-4}$. However, the expected early
  brightness is likely different from those of SN~2011fe and SN~2014J,
  since materials of our binary spread out to $\sim 0.1R_\odot$, which
  does not fit to what were inferred for these SNe.  Moreover, the
  explosion likely results in an SN remnant with an extremely
  asymmetric symmetric shape, which is unusual for SN~Ia.
\item In the carbon-ignited violent merger model, our binary can
  explode. However, the explosion likely results in the early
  brightness dissimilar to those of SN~2011fe and SN~2014J for the
  same reason as for the helium-ignited violent merger
  model. Moreover, we predict that the explosion will synthesize \fsni
  whose distribution is extremely aspherical. This is inconsistent
  with \fsni distribution generally inferred for SN~Ia. Note that
  \fsni distribution will depend on our choice of the raw and smoothed
  temperatures for the $^{12}$C + $^{12}$C reaction. In the case of
  the raw temperature, \fsni is absent in the center of the
  explosion. On the other hand, in the case of the smoothed
  temperature, \fsni distribution is similar to a hourglass.
\item If our binary explodes a few hundred seconds after its merger by
  some mechanism (i.e., the `other model'), the explosion may have the
  early brightness consistent with that of SN~2014J. At that time,
  materials of our binary spread out beyond $1R_\odot$.
\item For a particular set of the binary parameters examined in this
  paper, the binary would not lead to an SN Ia explosion through the
  Chandrasekhar mass model. Rather, the merger remnant should be
  converted to an oxygen-neon-magnesium WD, and then will experience
  an accretion-induced collapse to become a neutron star.
\end{itemize}

Binary CO~WDs generate the merger ejecta before and after its merger
time. The merger ejecta will interact with its surrounding ISM, and
form a merger shell. We estimate a bolometric luminosity of the merger
shell; the luminosity is more than $\sim 2 \times 10^{35}~\ergs$ at
its peak, if the total mass of the binary CO~WDs exceeds the
Chandrasekhar mass.  Suppose that all the super-Chandrasekhar binaries
explode in the Chandrasekhar mass model or fail to explode at all, the
number of the merger shells in Milky Way is estimate to be $\sim
10$. Detection of such merger shells can rule out the helium-ignited
and carbon-ignited violent merger models. If an explosion is initiated
in the Chandrasekhar mass model, a merger shell can be detected not
only from pre-explosion images of a site of an SN~Ia, but also in the
post-explosion observations. Unfortunately, we have not found merger
shells from pre-explosion images of previous SNe~Ia, since the lower
limit in these observations is at best $\sim 10^{36}~\ergs$. In
future, the merger shells would be found from post-explosion images of
sites of nearby SNe~Ia.

\section*{Acknowledgements}

Numerical simulations have been performed with HA-PACS at the Center
for Computational Sciences in University of Tsukuba. This research has
been supported in part by Grants-in-Aid for Scientific Research
(23224004, 23540262, 23740141, 24540227, 26400222, and 26800100) from
the Japan Society for the Promotion of Science, by World Premier
International Research Center Initiative (WPI Initiative), MEXT,
Japan, and by MEXT program for the Development and Improvement for the
Next Generation Ultra High-Speed Computer System under its Subsidies
for Operating the Specific Advanced Large Research
Facilities.

\appendix

\section{Artificial viscosity}
\label{sec:viscosity}

We describe our chosen artificial viscosity, and its numerical
parameters. In this section, variables are defined again with the same
notations as those in the main text. The viscosity exerted on
$i$-particle by $j$-particle is indicated as $\Pi_{ij}$. As described
in the main text, we adopt a sheer-free viscosity term
\citep{Balsara95} combined with time dependent viscosity parameters
\citep{Morris97}. Then, the term is given by
\begin{align}
  \Pi_{ij} = \bar{f}_{ij} \hat{\Pi}_{ij,{\rm max}},
\end{align}
where $\bar{f}_{ij}$ is so-called Balsara switch, and
$\hat{\Pi}_{ij,{\max}}$ is a sort of bulk and von-Neumann-Richtmyer
viscosities. Hereafter, we define
a combination of the overline and subscript ``$ij$'' as
\begin{align}
  \bar{X}_{ij} = \frac{1}{2} (X_i + X_j). \label{eq:bardef}
\end{align}

The viscosity term $\hat{\Pi}_{ij,{\max}}$ is expressed as
\begin{align}
  &\hat{\Pi}_{ij,{\rm max}} = \max \left( \hat{\Pi}_{ij,i},
  \hat{\Pi}_{ij,j} \right), \\
  &\hat{\Pi}_{ij,k} = \left\{
  \begin{array}{ll}
     \displaystyle \frac{- \alpha_k \bar{c}_{{\rm s},ij} \mu_{ij} +
       \beta_k \mu_{ij}^2}{\bar{\rho}_{ij}} & (\bm{r}_{ij} \cdot
     \bm{v}_{ij} < 0) \\ 
     0 & (\bm{r}_{ij} \cdot \bm{v}_{ij} \ge 0)
  \end{array}
  \right.,
\end{align}
where $\rho_i$ and $c_{{\rm s},i}$ are the mass density and sound
speed of $i$-particle, respectively. The vectors $\bm{r}_i$ and
$\bm{v}_i$ are, respectively, the position and velocity of
$i$-particle, and $\bm{r}_{ij}=\bm{r}_j-\bm{r}_i$ and
$\bm{v}_{ij}=\bm{v}_j-\bm{v}_i$. The variable $\mu_{ij}$ is given by
\begin{align}
  \mu_{ij} = \frac{\bar{h}_{ij} \bm{r}_{ij} \cdot
    \bm{v}_{ij}}{|\bm{r}_{ij}|^2 + \delta_1 \bar{h}_{ij}},
\end{align}
where $h_i$ is the kernel length of $i$-particle, and $\delta_1=0.01$
is chosen. The viscosity parameter $\alpha_i$ is time-dependent. It is
evolved as
\begin{align}
  \dot{\alpha}_i = - \frac{\alpha_i - \alpha_{\rm min}}{h_i/(\xi
    c_{{\rm s},i})} + \max \left[ - (\nabla \cdot \bm{v}_i)
    (\alpha_{\rm max} - \alpha_i), 0 \right], \label{eq:dotalpha}
\end{align}
where $\alpha_{\rm max}=1.5$, $\alpha_{\rm min}=0.05$, and
$\xi=0.25$. Another viscosity parameter $\beta_i$ is proportional to
$\alpha_i$, such that $\beta_i = 2 \alpha_i$.

The Balsara switch can be
written by $f_i$ and $f_j$ with equation~(\ref{eq:bardef}), and $f_i$
is given by
\begin{align}
  f_{i} = \frac{|\nabla \cdot \bm{v}_i|}{|\nabla \cdot \bm{v}_i| +
    |\nabla \times \bm{v}_i| + \delta_2 c_{{\rm s},i} /
    h_i}, \label{eq:balsara}
\end{align}
where we adopt $\delta_2=0.0001$.

\clearpage

\label{lastpage}

\end{document}